\documentclass[twocolumn,nofootinbib,superscriptaddress,byrevtex,showpacs]{revtex4}
\usepackage{amsmath,amssymb,amsbsy,mathbbol}
\usepackage{graphicx,subfigure}
\usepackage{comment}

\def\str{{\mathrm{str}}}

\begin{document}

\unitlength=1mm

% Greek Letters
\def\a{{\alpha}}
\def\b{{\beta}}
\def\d{{\delta}}
\def\D{{\Delta}}
\def\e{{\epsilon}}
\def\g{{\gamma}}
\def\G{{\Gamma}}
\def\k{{\kappa}}
\def\l{{\lambda}}
\def\L{{\Lambda}}
\def\m{{\mu}}
\def\n{{\nu}}
\def\w{{\omega}}
\def\O{{\Omega}}
\def\S{{\Sigma}}
\def\s{{\sigma}}
\def\t{{\tau}}
\def\th{{\theta}}
\def\x{{\xi}}

\def\ol#1{{\overline{#1}}}

%slash's
\def\Dslash{D\hskip-0.65em /}
\def\dslash{{\partial\hskip-0.5em /}}
\def\vslash{{\rlap \slash v}}
\def\qbar{{\overline q}}

% Jargon
\def\CPT{{$\chi$PT}}
\def\QCPT{{Q$\chi$PT}}
\def\PQCPT{{PQ$\chi$PT}}
\def\tr{\text{tr}}
\def\str{\text{str}}
\def\diag{\text{diag}}
\def\order{{\mathcal O}}
\def\vit{{\it v}}
\def\vD{\vit\cdot D}
\def\am{\alpha_M}
\def\bm{\beta_M}
\def\gm{\gamma_M}
\def\smb{\sigma_M}
\def\smt{\overline{\sigma}_M}
\def\tb{{\tilde b}}

\def\mc#1{{\mathcal #1}}

% Fields
\def\Bbar{\overline{B}}
\def\Tbar{\overline{T}}
\def\cBbar{\overline{\cal B}}
\def\cTbar{\overline{\cal T}}
\def\pq{(PQ)}

\def\eqref#1{{(\ref{#1})}}

%\preprint{CALT 68-2615}

%\preprint{DOE/ER/40762-374}

%\preprint{NT@UW 06-18}

%\vphantom{}

\title{\bf Two Meson Systems with Ginsparg-Wilson Valence Quarks}

\author{Jiunn-Wei Chen}
\email[]{jwc@phys.ntu.edu.tw} 
\affiliation{Department of Physics and Center for Theoretical Sciences, National Taiwan University, Taipei 10617, Taiwan}

\author{Donal O'Connell}
\email[]{donal@theory.caltech.edu}
\affiliation{California Institute of Technology, Pasadena, CA 91125, USA}

\author{ Andr\'e Walker-Loud} 
\email[]{walkloud@umd.edu}
\affiliation{Department of Physics, University of Maryland,
	College Park, MD 20742-4111}
\affiliation{Department of Physics, University of Washington,
	Box 351560,
	Seattle, WA 98195-1560, USA}

%\date{\mydate}
\begin{abstract}
Unphysical effects associated with finite lattice spacing and
partial quenching may lead to the presence of unphysical terms in
chiral extrapolation formulae. These unphysical terms must then be
removed during data analysis before physical predictions can be made.
In this work, we show that through next-to-leading order, there are
no unphysical counterterms in the extrapolation formulae, expressed in
lattice-physical parameters, for meson scattering lengths in theories
with Ginsparg-Wilson valence quarks. Our work applies to 
most sea quark discretization, provided that chiral perturbation theory
is a valid approximation.  We demonstrate our results with explicit
computations and show that, in favorable circumstances, the extrapolation
formulae do not depend on the unknown constant $C_\mathrm{Mix}$ appearing
at lowest order in the mixed action chiral Lagrangian.  We show that the
$I=1\ KK$ scattering length does not depend on $C_\mathrm{Mix}$ in contrast to
the $I=3/2\ K\pi$ scattering length.  In addition, we show that these observables
combined with $f_K / f_\pi$ and the $I=2$ $\pi\pi$ scattering length
share only two linearly independent sets of counterterms, providing a
means to test the mixed action theory at one lattice spacing.  We therefore make a prediction for the $I=1\ KK$ scattering length.
\end{abstract}

\pacs{12.38.Gc}
\maketitle

\newpage

%\pagebreak
%\tableofcontents
%\newpage

%%%%%%%%%%%%%%%%%%%%%%%%%%%%%%%%%%%%%%%%%%%%%%%%%%%
%
%	Introduction
%
%%%%%%%%%%%%%%%%%%%%%%%%%%%%%%%%%%%%%%%%%%%%%%%%%%%
\section{Introduction}

There is currently a tension in lattice simulations of QCD phenomena between the need for quarks obeying chiral symmetry on the lattice, and the need for quark masses light enough that one is in the chiral regime. This tension occurs because quark discretization schemes which obey chiral symmetry on the lattice, such as domain wall fermions~\cite{Kaplan:1992bt,Shamir:1993zy,Furman:1994ky} or overlap fermions~\cite{Narayanan:1993sk,Narayanan:1993ss,Narayanan:1994gw}, both of which satisfy the Ginsparg-Wilson relation~\cite{Ginsparg:1981bj,Luscher:1998pq}, are numerically expensive to simulate.  On the other hand, Wilson fermions~\cite{Wilson:1974sk} or staggered fermions~\cite{Kogut:1974ag,Susskind:1976jm} are faster but violate chiral symmetry at non-zero lattice spacing.

One way of resolving this tension is to recognize that the most
computationally intensive stage of a fully dynamical simulation
is the evaluation of the quark determinant. This determinant is
associated with the sea quarks and is a component of the probability
measure on the space of gauge field configurations. This observation
has long been the motivation for partial quenching (PQ)~\cite{Bernard:1993sv,Sharpe:1997by} --- sea quark masses are taken to be larger than valence quark masses so that the sea
quarks are more localized and the determinant is easier to compute.
The notion of a ``mixed action" (MA) simulation takes this line of
reasoning one step farther~\cite{Bar:2002nr,Bar:2003mh}. A mixed action simulation uses different
quark discretizations in the sea and valence sectors. In this case,
the valence quarks can be chosen to obey the Ginsparg-Wilson relation
so that they enjoy chiral symmetry at finite lattice spacing. The
numerically expensive sea quarks, on the other hand, can be chosen
to be inexpensive Wilson or (rooted) staggered sea quarks, for
example.

There have recently been a significant number of mixed action lattice QCD simulations, 
~\cite{Bowler:2004hs,Bonnet:2004fr,Edwards:2005ym,Beane:2005rj,Beane:2006mx,Beane:2006pt,Beane:2006fk,Beane:2006kx,Alexandrou:2006mc,Beane:2006gj,Bar:2006zj}, the majority of which have employed domain wall valence fermions on the publicly available MILC lattices~\cite{Bernard:2001av}.%
\footnote{The MILC lattices themselves utilize asqtad-improved~\cite{Orginos:1998ue,Orginos:1999cr}, staggered sea fermions.} %
The effective theories appropriate for mixed action simulations were originally developed in Refs.~\cite{Bar:2002nr,Bar:2003mh} for Ginsparg-Wilson (GW) valence fermions on Wilson sea fermions, and later for GW valence fermions on staggered sea fermions~\cite{Bar:2005tu}; these theories have received considerable theoretical attention recently~\cite{Tiburzi:2005vy,Golterman:2005xa,Tiburzi:2005is,Chen:2005ab,Bunton:2006va,Aubin:2006hg} in response to the numerical interest. 

In this article, we study aspects of the chiral perturbation theories
appropriate for mesonic processes in mixed action simulations with
Ginsparg-Wilson valence quarks. The chiral properties of the
Ginsparg-Wilson valence quarks are central to our work, so we adopt
the convention that when we refer to a mixed action simulation, we
imply that the valence quarks are Ginsparg-Wilson.
We work consistently at next-to-leading order (NLO) in the effective
field theory expansion, which is a dual expansion in powers of the quark mass, $m_q$ and the lattice spacing, $a$.  At this order, one can view current lattice
simulations as being methods of computing the values of certain
coefficients in the NLO chiral Lagrangian, which for mesonic quantities is known as the Gasser-Leutwyler Lagrangian~\cite{Gasser:1983yg,Gasser:1984gg}.  This is because lattice simulations are
performed at quark masses larger than the physical quark mass, so
that lattice data must be fit to formulae computed in chiral
perturbation theory. These fits determine the unknown coefficients
occurring in the chiral formulae, known as low energy constants (LEC)s, 
so that the chiral expression can then be used at the physical
values of the meson masses and decay constants to predict the results
of physical experiments.  Frequently there are non-physical operators in the NLO chiral Lagrangians
describing discretized fermions. For example, there are of order
100 operators in the NLO staggered chiral Lagrangian~\cite{Sharpe:2004is}
compared to order 10 in the Gasser-Leutwyler Lagrangian.  These
unphysical operators lead to unphysical terms in chiral extrapolation
formulae which must somehow be removed to make physical predictions.
One might think that this will also be an issue in mixed action
chiral perturbation theory (MA$\chi$PT), since there are certainly
many additional operators at NLO.  However, we show that mixed
action lattice simulations of mesonic scattering lengths do not
depend on any unphysical operators at NLO, if these scattering
lengths are expressed in terms of the pion mass measured on the
lattice and the decay constant measured on the lattice~\cite{O'Connell:2006sh}.  For linguistic brevity, we will refer to the pion mass measured on the lattice as the lattice-physical pion mass and similarly for the decay constant.

Each choice of sea quark discretization leads to a different
MA$\chi$PT.  For example, tree level shifts of the masses of mesons
composed of two sea quarks, or of one valence and one sea quark, are
different for staggered sea quarks and for Wilson sea quarks.
Therefore, one may think that the chiral extrapolation formulae
depend on the nature of the sea quark discretization. We show that,
at NLO, the only difference between the extrapolation formulae is
in the leading order mass shifts of the mesons composed of two sea quarks.  Thus, once these mass shifts are known, one can use the same extrapolation formulae for different
sea quark discretizations.%
\footnote{For staggered fermions, these mass corrections to the mesons are well know~\cite{Aubin:2004fs}.  However, these effects are less well determined for Wilson fermions.} 
In fact, any sea quark
discretization will do provided,%
\footnote{In this paper, we restrict ourselves to isospin symmetric masses in the valence and sea sectors. } 
firstly, that QCD is recovered in
the continuum limit, and secondly, that the sea-sea mesons may be
described at leading order by chiral perturbation theory with the usual kinetic and
mass terms, or that the non-locality of the appropriate chiral perturbation theory is correctly captured by
the replica method~\cite{Damgaard:2000gh}.%
\footnote{We have in mind the current discussion regarding whether rooted staggered fermions become QCD in the continuum limit.  There is growing numerical and formal evidence lending support to the hope that rooted staggered fermions are in the same universality class as QCD.  We refer the reader to Ref.~\cite{Sharpe:2006re} for current summary of the issues.} 
In addition, we have assumed that the quarks of the
sea sector are only distinguished by their masses, so that, for example, the
same discretization scheme has been used for all sea quarks, and we assume
that chiral perturbation theory itself is a valid approximation.

There are still various challenges facing mixed action simulations.  
Mixed action simulations always violate unitarity at finite lattice spacing.%
\footnote{For arbitrarily small lattice spacings, the differences arising from the different lattice actions will become negligible, which practically means smaller than the statistical and systematic uncertainties for a given observable, and these unitarity violating terms will no longer be important (assuming the quark masses are tuned equal).  This of course, also implies that a MA effective field theory description will no longer be necessary, however for MA lattice simulations today and the foreseeable future, MA$\chi$PT is the necessary tool for controlling extrapolations to the physical point.} 
In MA$\chi$PT and PQ$\chi$PT, the most severe unitarity violations are encoded in hairpin propagators of flavor neutral mesons.  We point out a simple parametrization which allows a convenient book-keeping of these unitarity violating effects.  Additionally, the value of the new constant, $C_\mathrm{Mix}$, that appears in the LO mixed action Lagrangian, is currently unknown.  This term leads to an additive lattice spacing dependent mass shift of ``mixed" mesons consisting of one valence and one sea quark.  This  causes a mismatch of the meson masses composed of different quarks but does not play a role in the well known enhanced chiral logarithms~\cite{Bernard:1992mk,Sharpe:1992ft,Sharpe:1997by} or the enhanced power-law volume dependence of two-hadron states~\cite{Bernard:1995ez}.%
\footnote{In this article we do not discuss the observed negative norm issues involving scalar meson correlators~\cite{Aubin:2004fs,Gregory:2005yr}, but these can at least be qualitatively understood with the appropriate effective field theory methods~\cite{Prelovsek:2005rf,Bernard:2006gj}.} 
In addition, the value of this constant is presumably different for each sea quark discretization.  However, we show that under favorable circumstances physical quantities such as scattering lengths do not depend on this constant, as was first noticed in Ref.~\cite{Chen:2005ab}.  

Finally, to demonstrate our arguments, we determine various NLO formulae for use in chiral extrapolation of certain mesonic quantities. We have computed the $KK$ and $K\pi$
scattering lengths in $SU(6|3)$ MA$\chi$PT.  In addition, for completeness, we present the $\pi \pi$ scattering length in $SU(4|2)$ and $SU(6|3)$ MA$\chi$PT, which were computed in Ref.~\cite{Chen:2005ab} as well as the $\pi$ and $K$ meson masses and decay constants, which were first computed in Refs.~\cite{Bar:2002nr,Bar:2003mh,Bar:2005tu}, but we express these quantities in terms of the PQ parameters we introduce in Sec.~\ref{sec:LO}.  We conclude with a discussion of our results, and in particular show that among the three meson scattering lengths mentioned above and the quantity $f_K / f_\pi$ there are only two linearly independent counterterms at NLO, which are the corresponding physical counterterms of $\chi$PT.  Therefore, these four processes provide a means to test the MA formalism with only one lattice spacing.  In the appendices we collect the various formulae which are necessary for the chiral extrapolations of the quantities we discuss in this paper.

%%%%%%%%%%%%%%%%%%%%%%%%%%%%%%%%%%%%%%%%%%%%%%%%%%%
%
%				MA EFT
%
%%%%%%%%%%%%%%%%%%%%%%%%%%%%%%%%%%%%%%%%%%%%%%%%%%%
\section{Mixed Action Effective Field Theory \label{sec:MAEFT}}

We will not give a thorough introduction to mixed action or partially quenched theories here.  We will simply give a brief review to introduce our notation and power counting.  For a good introduction to MA theories we refer the reader to Refs.~\cite{Bar:2002nr,Bar:2003mh,Bar:2005tu}, and for PQ theories to Refs.~\cite{Sharpe:2000bc,Sharpe:2001fh}.

%%%%%%%%%%%%%%%%%%%%%%%%%%%%%%%%%%%%%%%%%%%%%%%%%%%
%
%		MA at LO
%
%%%%%%%%%%%%%%%%%%%%%%%%%%%%%%%%%%%%%%%%%%%%%%%%%%%
\subsection{Mixed Actions at Lowest Order \label{sec:LO}}

To construct the appropriate Lagrangian to a given order, one must specify a power counting.  As mentioned above, $\chi$PT is a systematic expansion about the zero momentum, zero quark mass limit, for which the small expansion parameter is
\begin{equation}
	\varepsilon_m^2 \sim \frac{p^2}{\Lambda_\chi^2} \sim \frac{m_\pi^2}{\Lambda_\chi^2}  ,
\end{equation}
where $m_\pi^2 \propto m_q$.  For effective theories extended to include lattice spacing artifacts, one must include an additional small parameter.%
\footnote{The general procedure~\cite{Sharpe:1998xm} is to construct the continuum Symanzik quark level effective theory for a given lattice action~\cite{Symanzik:1983gh,Symanzik:1983dc} and then build the low energy effective theory with spurion analysis on this continuum lattice action.} 
We will be interested in theories for which the leading sea quark lattice spacing dependence is $\mc{O}(a^2)$, such as staggered, $\mc{O}(a)$-improved Wilson~\cite{Sheikholeslami:1985ij}, twisted-mass at maximal twist~\cite{Frezzotti:2000nk}, or chiral fermions.  Therefore, we shall denote the small parameter counting lattice spacing artifacts to be 
\begin{equation}
	\varepsilon_a^2 \sim a^2\, \Lambda_{QCD}^2\, ,
\end{equation}
and we shall work consistently in the dual expansion to 
\begin{equation}\label{eq:power_counting}
	\mc{O}(\varepsilon_m^4)\quad ,\quad \mc{O}(\varepsilon_m^2\, \varepsilon_a^2)\quad ,\quad 
	\mc{O}(\varepsilon_a^4)\, .
\end{equation}

At leading order (LO) in the quark mass expansion, the mixed action Lagrangian is simply given by the partially quenched Lagrangian~\cite{Bar:2002nr},
\begin{equation}\label{eq:PQChPT}
	\mc{L} = \frac{f^2}{8} \str \left( \partial_\mu \Sigma \partial^\mu \Sigma^\dagger \right)
		+ \frac{f^2 B_0}{4} \str \left( m_q \Sigma^\dagger + \Sigma m_q^\dagger \right)\, ,
\end{equation}
where we use the normalization $f \simeq 132$~MeV, and
\begin{equation}
	\S = {\rm exp} \left( \frac{2 i \Phi}{f} \right) ,  \;\;\; 
	\Phi = \begin{pmatrix}
			M & \chi^\dagger\\
			\chi & \tilde{M}\\
		\end{pmatrix}\, .
\label{eq:sigma}
\end{equation}
The matrices $M$ and $\tilde{M}$ contain bosonic mesons while $\chi$ and $\chi^\dagger$ contain fermionic mesons with one ghost quark or antiquark.  To be specific, we will discuss the theory with 3 valence (and ghost) quarks and 3 sea quarks, for which%
\footnote{For staggered sea quarks, each sea quark label implicitly includes a taste label as well.  For example, $\phi_{uj}$ is a $1\times4$ vector in taste-space.} 
%
%\begingroup
%\small
\begin{align}\label{eq:mesons}
	M =\begin{pmatrix}
			\eta_u & \pi^+ & K^+ & \phi_{uj} & \phi_{ul} & \phi_{ur} \\
			\pi^- & \eta_d & K^0 & \phi_{dj} & \phi_{dl} & \phi_{dr} \\
			K^- & \ol{K}^0 & \eta_s & \phi_{sj} & \phi_{sl} & \phi_{sr} \\
			\phi_{ju} & \phi_{jd} & \phi_{js} & \eta_j & \phi_{jl} & \phi_{jr} \\
			\phi_{lu} & \phi_{ld} & \phi_{ls} & \phi_{lj} & \eta_l & \phi_{lr} \\
			\phi_{ru} & \phi_{rd} & \phi_{rs} & \phi_{rj} & \phi_{rl} & \eta_{r} \\
		\end{pmatrix}\quad ,\quad 
\nonumber\\
	\tilde M = \begin{pmatrix}
			{\tilde \eta}_u & {\tilde \pi}^+ & \tilde{K}^+ \\
			{\tilde \pi}^- & {\tilde \eta}_d & \tilde{K}^0 \\
			\tilde{K}^- & \tilde{\ol{K}}^0 & \tilde{\eta}_{s} \\
		\end{pmatrix}\quad ,\quad
%	\nonumber\\
	\chi^\dagger = \begin{pmatrix}
		\phi_{u \tilde{u}} & \phi_{u \tilde{d}} & \phi_{u \tilde{s}} \\
		\phi_{d \tilde{u}} & \phi_{d \tilde{d}} & \phi_{d \tilde{s}} \\
		\phi_{s \tilde{u}} & \phi_{s \tilde{d}} & \phi_{s \tilde{s}} \\
		\phi_{j \tilde{u}} & \phi_{j \tilde{d}} & \phi_{j \tilde{s}} \\
		\phi_{l \tilde{u}} & \phi_{l \tilde{d}} & \phi_{l \tilde{s}} \\
		\phi_{r \tilde{u}} & \phi_{r \tilde{d}} & \phi_{r \tilde{s}} \\
		\end{pmatrix}.
\end{align}
%\endgroup
%
The upper $N_v \times N_v$ block of $M$
contains the usual mesons composed of a valence quark and anti-quark.  The lower $N_s \times N_s$ block of $M$ contains the sea quark-antiquark mesons and the off-diagonal block elements of $M$ contain bosonic mesons of mixed valence-sea type.  

For MA theories there are two types of operators we need to consider at LO in $\varepsilon_a^2$.  There are those which modify the sea-sea sector meson potential, which we shall denote as $\mc{U}_{sea}$, and those which modify the mixed meson potential, which we shall denote as $\mc{U}_{VS}$, such that the Lagrangian, Eq.~\eqref{eq:PQChPT} is modified by the additional terms (following the sign conventions of Ref.~\cite{Bar:2005tu}),
\begin{equation}\label{eq:MA_potential}
	\mc{L}_{MA} = -a^2 \Big( \mc{U}_{sea} -\mc{U}_{VS} \Big)\, .
\end{equation}
We shall not specify the form of the sea-sea meson potential, $\mc{U}_{sea}$, but only note again that at the order we are concerned with, we only need to know how the masses of the sea-sea mesons are modified at LO in $\varepsilon_a^2$ which have been discussed, for example, in Refs.~\cite{Bar:2002nr,Bar:2005tu}.  The other important thing to know is that the structure of $\mc{U}_{VS}$ is independent of the type of sea quark and is given by~\cite{Bar:2002nr,Bar:2005tu}
\begin{equation}\label{eq:MALO_opp}
	\mc{U}_{VS} = C_\mathrm{Mix}\, \str \Big( T_3\, \Sigma\, T_3\, \Sigma^\dagger \Big)\, ,
\end{equation}
where the flavor matrix $T_3$ is a difference in projectors onto the valence and sea sectors of the theory,
\begin{equation}
	T_3 = \mc{P}_S -\mc{P}_V = \textrm{diag} (-I_V, I_S, -I_V)\, .
\end{equation}
This operator leads to an additive shift of the valence-sea meson masses, such that all the pseudo-Goldstone mesons composed of either valence quarks, $v$, sea quarks, $s$, or both have LO masses given by%
\footnote{Here and throughout this article, we use \textit{tilde}s over the masses to indicate additive lattice spacing corrections to the meson masses.} 
\begin{align}\label{eq:LOMasses}
	m_{v_1 v_2}^2 &= B_0 (m_{v_1} +m_{v_2} )\, ,\nonumber\\
	\tilde{m}_{vs}^2 &= B_0 (m_v +m_s ) +a^2 \Delta_\mathrm{Mix}\, ,\nonumber\\
	\tilde{m}_{s_1 s_2}^2 &= B_0 (m_{s_1} +m_{s_2} ) +a^2 \Delta_{sea}\, ,
\end{align}
with $\Delta_{sea}$ determined by $\mc{U}_{sea}$ and%
\footnote{For a twisted mass sea~\cite{Frezzotti:2000nk}, one must keep separate track of the neutral and charged mesons as they receive a relative $\mc{O}(a^2)$ splitting~\cite{Sharpe:2004ny}, similar to the various \textit{taste} mesons for staggered fermions.}
\begin{equation}
	\Delta_\mathrm{Mix} = \frac{16\, C_\mathrm{Mix}}{f^2}\, .
\end{equation}

The most severe and well known unitarity violating feature of PQ and MA theories is the presence of double pole propagators in the flavor neutral mesons~\cite{Bernard:1993sv}.  In particular the momentum space propagators between two mesons composed of valence quarks of flavors $a$ and $b$ respectively are given by%
\footnote{In mixed action theories, there are additional hairpin interactions proportional to the lattice spacing which arise from unphysical operators in the theory, similar to the lattice spacing dependent hairpin interactions in staggered $\chi$PT~\cite{Aubin:2003mg}.  For $\mc{O}(a)$ improved Wilson fermions and staggered fermions, these effects are higher order than we are concerned with in this paper~\cite{Golterman:2005xa}, which is also true for twisted mass fermions at maximal twist.  For Wilson and twisted mass fermions (away from maximal twist), these effects appear at the order we are working, and must be included.  We assume for the rest of the paper that the sea quark scaling violations are $\mc{O}(a^2)$ or higher.}

\begin{align}\label{eq:etaPropSU63}
\mc{G}_{ab}(p^2) =&\ \frac{i \d_{ab}}{p^2 - m_{aa}^2}
	\nonumber\\& \quad
	-\frac{i}{3} \frac{ (p^2 -\tilde{m}_{jj}^2) (p^2 - \tilde{m}_{rr}^2)}
		{(p^2 -m_{aa}^2)(p^2 -m_{bb}^2) (p^2 -\tilde{m}_X^2)}
\end{align}
where $\tilde{m}_X$ is the mass of the $\eta_{sea}$-field,
\begin{align}\label{eq:etaMass}
	\tilde{m}_X^2 &= \frac{1}{3}\tilde{m}_{jj}^2 +\frac{2}{3}\tilde{m}_{rr}^2\, .
\end{align}
When the valence quark masses are equal, either in the isospin limit of light quarks or for the same flavor, $a=b$, the above propagator acquires a double pole.  It is these double poles which lead to the well-known sicknesses of PQ$\chi$PT, such as the enhanced chiral logs~\cite{Bernard:1992mk,Sharpe:1992ft,Sharpe:1997by}, and enhanced power-law volume dependence of two-particle states~\cite{Bernard:1995ez}. Here, we introduce what we call ``partial quenching parameters," which are a difference in the quark masses for PQ theories and, more generally for MA theories, a difference in the masses of mesons composed of two sea quarks and two valence quarks.  For PQ theories, when these quantities are zero, the theory reduces to an unquenched theory. For MA theories, when one tunes these parameters to zero, one tunes the double pole structure of the flavor-neutral meson propagators to zero up to higher order corrections, and thus has the most QCD-like scenario for a MA theory (we note that there is still a mismatch in the mass of the mixed meson, composed of one valence and one sea quark, Eq.~\eqref{eq:LOMasses}, from the others).  We therefore introduce the partial quenching parameters,%
\footnote{For a staggered sea it is the taste-identity meson masses which enter these PQ parameters~\cite{Bar:2005tu} (which have been measured~\cite{Aubin:2004fs}) while for a twisted mass sea, it is the neutral pion mass.} 
\begin{align}\label{eq:PQparameters}
	\tilde{\D}_{ju}^2 &\equiv \tilde{m}_{jj}^2 - m_{uu}^2
		= 2 B_0 (m_j- m_u) + a^2 \Delta_{sea} +\dots\, , \nonumber\\
	\tilde{\D}_{rs}^2 &\equiv \tilde{m}_{rr}^2 - m_{ss}^2
		= 2 B_0 (m_r- m_s) + a^2 \Delta_{sea} +\dots\, ,
\end{align}
where the dots denote higher order corrections to the meson masses.  We will now move on to discuss the general structure of mixed action theories for arbitrary sea quarks at the next order.

%%%%%%%%%%%%%%%%%%%%%%%%%%%%%%%%%%%%%%%%%%%%%%%%%%%
%
%	MA ChPT
%
%%%%%%%%%%%%%%%%%%%%%%%%%%%%%%%%%%%%%%%%%%%%%%%%%%%
\subsection{Mixed Action $\chi$PT at NLO}\label{sec:NLO}

It was shown in Ref.~\cite{Chen:2005ab} that the $I=2$  $\pi\pi$ scattering length at NLO, expressed in terms of the bare parameters of the chiral Lagrangian, is%
\footnote{Here, we show the scattering length for a two-sea flavor theory.  In the appendix, we also list the result for the three-sea flavor theory.  However, the following discussion of the counterterm structure of the NLO Lagrangian is independent of the number of sea flavors.}
%
%\begin{widetext}
\begin{multline}
       m_{\pi} a_{\pi\pi}^{I=2} = -\frac{m_{uu}^2}{8 \pi f^2} \Bigg\{ 1 
                +\frac{m_{uu}^2}{(4\pi f)^2} \Bigg[
                        4 \ln \left( \frac{m_{uu}^2}{\mu^2} \right) \\
                +4 \frac{\tilde{m}_{ju}^2}{m_{uu}^2} \ln \left( \frac{\tilde{m}_{ju}^2}{\mu^2} \right) 
                -1 +\ell^\prime_{\pi\pi}(\mu) \Bigg]
                \\
		- \frac{m_{uu}^2}{(4\pi f)^2} \Bigg[
             	   	\frac{\tilde{\D}_{ju}^4}{6 m_{uu}^4}
			+\frac{\tilde{\D}_{ju}^2}{m_{uu}^2} \left[ \ln \left( \frac{m_{uu}^2}{\mu^2} \right) +1 \right]
		\Bigg] 
	\\ 
                + \frac{\tilde{\D}_{ju}^2}{(4\pi f)^2}\, \ell^\prime_{PQ}(\mu) 
                + \frac{a^2}{(4\pi f)^2} \ell^\prime_{a^2}(\mu)
                \Bigg\}.
\label{eq:2seaBare}
\end{multline}
%\end{widetext}
Let us point out some features of this expression which are relevant from the point of view of chiral extrapolations.  Equation~\eqref{eq:2seaBare} depends on the mass $\tilde m_{ju}$ of a mixed valence-sea meson and consequently the expression depends on the value of the parameter $C_\mathrm{Mix}$. 
In addition, there is a dependence on the unphysical unknowns $\ell^\prime_{PQ}(\mu)$ and $\ell^\prime_{a^2}(\mu)$, as well as the decay constant and chiral condensate in the chiral limit, $f$ and $B_0$. Thus, Eq.~\eqref{eq:2seaBare} depends on three unphysical unknown parameters and three physical unknown parameters, $\ell^\prime_{\pi\pi}(\mu), f$ and $B_0$. One must fit all unknown parameters to extrapolate lattice data but only three are of intrinsic interest.

In terms of lattice-physical parameters, the same scattering length becomes
%\begin{widetext}
\begin{multline}\label{eq:2seaScattLength}
	m_\pi a_{\pi\pi}^{I=2}= -\frac{m_\pi^2}{8 \pi f_\pi^2} \Bigg\{ 1 
                + \frac{m_\pi^2}{(4\pi f_\pi)^2} \bigg[ 
		3\ln \left( \frac{m_\pi^2}{\mu^2} \right) 
	\\
	-1 
	-l_{\pi\pi}^{I=2}(\mu) 
	-\frac{\tilde{\D}_{ju}^4}{6\, m_\pi^4} 
	\bigg] \Bigg\}\, .
\end{multline}
%\end{widetext}
Notice that this expression does not depend on the mixed valence-sea mesons, and, in fact, the only unknown terms in the expression is the physical parameter $\ell_{\pi\pi}(\mu)$, and the sea-sea meson mass shift in $\tilde{\D}_{ju}^2$, Eq.~\eqref{eq:LOMasses}, which is already determined for staggered sea-quarks. Thus, chiral extrapolations using the formula Eq.~\eqref{eq:2seaScattLength} requires fitting only one parameter (two for non-staggered sea quarks), in contrast to chiral extrapolations using the scattering length expressed in terms of the bare parameters, Eq.~\eqref{eq:2seaBare}. Our goal in this section is to understand the origin of this simplification, and under what circumstances we may expect similar simplifications to occur in other processes. To do so, we must discuss the structure of the NLO terms of the MA$\chi$PT Lagrangian.

The symmetry structure of the underlying mixed action form of QCD determines the NLO operators in the mixed action chiral Lagrangian through a spurion analysis. However, the symmetries enjoyed by the valence quarks are different to the symmetries of the sea quarks in a mixed action theory. In particular, we only consider GW valence quarks which have a chiral symmetry; the numerically cheaper sea quarks typically violate chiral symmetry. Thus, it is helpful to consider spurions arising from the valence sector separately to the spurions of the sea sector.

The valence sector only violates chiral symmetry explicitly through the quark mass. Therefore, at NLO,%
\footnote{Lattice artifacts such as Lorentz symmetry violation lead to the presence of unphysical operators in the chiral Lagrangian. These operators will not be important in the following, as they are higher order in the chiral expansion for mesons~\cite{Bar:2003mh} (however they are relevant at $\mc{O}(\varepsilon_a^2)$ for baryons~\cite{Tiburzi:2005vy}).} %
the purely valence spurions are identical to the spurions in continuum, unquenched chiral perturbation theory, and so the valence-valence sector of the NLO mixed action chiral Lagrangian is the Gasser-Leutwyler Lagrangian.  The sea sector is different. At finite lattice spacing, the sea sector has enhanced sources of chiral symmetry violation --- for example, there are additional spurions associated with taste violation if the sea quarks are staggered, or in the case of a Wilson sea, the Wilson term violates chiral symmetry. Consequently, there are additional spurions in the sea sector. Of course, these spurions must involve the sea quarks and must vanish when
the sea quark fields vanish.%
\footnote{Not all the lattice spacing dependence may be captured with spurion analysis.  There are $\mc{O}(a^2)$  operators at the quark level which do not break chiral symmetry, for example, $\mc{O}^{(6)} = a^2\, \ol{Q}\, \Dslash^3\, Q$. This operator leads to an $a^2$ renormalization of all the low energy constants (LEC)s of the low energy theory.
% (and is present regardless of the lattice action, except for a perfect action), and thus does not modify our following arguments.  
Because this operator does not break any of the continuum QCD symmetries, it can not be distinguished through spurion analysis~\cite{Bar:2003mh}. Of course, an operator of this form is present when the QCD Lagrangian is run from a high scale (say, the weak scale) down to the scale of the lattice, so its effects could in principle by accounted for by performing a perturbative matching computation between the QCD effective Lagrangian at the scale $\mu = a^{-1}$ and the lattice action.
} 
Nevertheless, scattering amplitudes expressed in terms of lattice-physical parameters do not explicitly depend on the lattice spacing $a$, as we will now discuss.

In this paper, we work consistently to NLO in the MA$\chi$PT power
counting which we have defined in Eq.~\eqref{eq:power_counting}. At this order, the NLO operators in the Lagrangian are only used as counterterms; that is, at NLO one only computes at
tree level with the NLO operators. Since the in/out states used in
lattice simulations involve purely valence quarks, we must project
the NLO operators onto the purely valence quark sector of the theory.
Consequently, all of the spurions which involve the sea quark fields
vanish. Since the remaining spurions involve the valence quarks
alone, we only encounter the symmetry structure of the valence
quarks as far as the NLO operators are concerned. These spurions
only depend on quark masses and the quark condensate itself, and
so there can be no dependence on lattice discretization effects
arising in this way.  The exception to this argument arises in the case of double trace operators in the NLO chiral Lagrangian; in these cases the valence and sea sectors interact in a flavor-disconnected manner, unlike the operator in Eq.~\eqref{eq:MALO_opp}.  If one trace involves a valence-valence spurion while the other involves a sea-sea spurion, then the trace over the sea may still contribute to a physical quantity, for example the meson masses and decay constants, see App.~\ref{app:2flav_mpi_fpi}--\ref{app:decay_constants} for explicit examples. Note that the valence-valence operators which occur in these double trace operators must be proportional to one of the two operators present in the LO chiral Lagrangian, Eq.~\eqref{eq:PQChPT}.  Thus for meson scattering processes, the dependence upon the sea quarks from these double trace operators can only involve a renormalization of the leading order quantities $f$ and $B_0$.  Both the explicit sea quark mass dependence and the explicit lattice spacing dependence are removed from the scattering processes expressed in terms of the lattice-physical parameters since they are eliminated in favor of the decay constants and meson masses which can simply be measured on the lattice.  We therefore conclude that when expressed in lattice-physical parameters, there can be no dependence upon the sea quark masses leading to unphysical PQ counterterms and similarly there can be no dependence upon an unphysical lattice-spacing counterterm.  

Let us present another more physically intuitive argument concerning the absence of sea quark mass dependence in meson scattering processes. To do so, we must digress briefly on $\pi \pi$ scattering in $SU(3)$ chiral perturbation theory.  The strange quark mass $m_s$ is a
parameter of $SU(3)$ $\chi$PT, and so one would expect that the $\pi\pi$
scattering length includes analytic terms involving $m_s$.  However,
we can consider a theory in which the strange quark is heavy, so that we may integrate it out;
we must then recover $SU(2)$ $\chi$PT. Chiral symmetry forces any
$m_s$ dependence in the analytic terms of the $\pi \pi$ amplitude to
occur in the form $m_\pi^2 m_K^2$. But the only counterterm in
the on-shell $SU(2)$ scattering amplitude (Eq.~\eqref{eq:2seaScattLength}
with $\tilde{\Delta}_{ju} = 0$) is proportional to $m_\pi^4$. It is not
possible to absorb $m_\pi^2 m_K^2$ into $m_\pi^4$, so there can be no
$m_s$ dependence in the $SU(3)$ $\pi\pi$ scattering amplitude. This is
indeed the case, as was observed in Ref.~\cite{Knecht:1995tr}.

Now, let us return to the PQ and MA theories.  For the purposes of this discussion, we can ignore the flavor-neutral and ghost sectors, reducing our theory from an $SU(6|3)$ theory to
an $SU(6)$ theory. The sea quark dependence of this $SU(6)$ chiral
perturbation theory is analogous to the $m_s$ dependence of $SU(3)$
$\chi$PT in our $\pi\pi$ scattering example (as in this process the strange quark of $SU(3)$ only participates as a sea quark). A similar decoupling argument tells us that the sea quark
masses cannot affect processes involving the valence sector provided one
uses the analogues of on-shell parameters which are the lattice-physical
parameters.  We conclude that there can be no analytic dependence on the sea quark masses in a mesonic scattering amplitude.  Further, these arguments only depend upon the chiral symmetry of the valence quarks and thus also apply to the lattice spacing dependence.

Now, we shall make these arguments concrete by explicit computations.  The NLO Lagrangian describing the valence and sea quark mass dependence is the Gasser-Leutwyler Lagrangian with traces replaced by supertraces:
\begin{align}
\mathcal{L}_{GL} =&\ L_1 \left[ \str \left( \partial_\mu \Sigma \partial^\mu \Sigma^\dagger \right) \right]^2
	\nonumber\\&
	+ L_2 \; \str \left( \partial_\mu \Sigma \partial_\nu \Sigma^\dagger \right) \str \left( \partial^\mu \Sigma 		\partial^\nu \Sigma^\dagger \right) 
	\nonumber\\&
	+ L_3 \; \str \left( \partial_\mu \S \partial^\mu \S^\dagger \partial_\nu \S \partial^\nu \S^\dagger 
		\right) 
	\nonumber\\&
	+ 2 B_0\, L_4 \; \str \left( \partial_\mu \S \partial^\mu \S^\dagger \right) 
		\str \left ( m_q \S^\dagger + \S m_q^\dagger \right) 
	\nonumber\\&
	+ 2 B_0\, L_5 \; \str \left[ \partial_\mu \S \partial^\mu \S^\dagger 
		\left( m_q \S^\dagger + \S m_q^\dagger \right) \right] 
	\nonumber\\&
	+ 4 B_0^2\, L_6 \; \left[ \str \left( m_q \S^\dagger + \S m_q^\dagger \right) \right]^2 
	\nonumber\\&
	+ 4 B_0^2\, L_7 \; \left[ \str \left( m_q^\dagger \S - \S^\dagger m_q \right) \right]^2
	\nonumber\\&
	+ 4 B_0^2\, L_8 \; \str \left( m_q \S^\dagger m_q \S^\dagger 
		+ \S m_q^\dagger \S m_q^\dagger \right).
\label{eq:LGL}
\end{align}
Having a concrete expression for the Lagrangian,%
\footnote{The generators of the PQ and MA theories form graded groups and therefore lack the Cayley-Hamilton identities of $SU(N)$ theories.  Therefore, PQ and MA theories have additional operators compared to their $\chi$PT counterparts.  For example, the $\mc{O}(p^4)$ Lagrangian has one additional operator as compared to the Gasser-Leutwyler Lagrangian.  However, we do not need to consider the effects of this operator in our analysis as it has been shown that it can be constructed such that it does not contribute to valence quantities until $\mc{O}(p^6)$~\cite{Sharpe:2003vy}.  This is not generally the case, as is demonstrated by various examples in the baryon sector~\cite{Chen:2001yi,Beane:2002vq,Beane:2002np,Walker-Loud:2004hf,Tiburzi:2004rh,Detmold:2005pt,Tiburzi:2005na,O'Connell:2005un,Detmold:2006vu,Walker-Loud:2006sa}.} 
 we can easily show explicitly how the sea quark mass dependence disappears. The key is that when constructing NLO correlation functions of purely valence quarks, we can replace the mesonic matrix $\Phi$ in the NLO Lagrangian by a projected matrix
\begin{equation}
\Phi \rightarrow P_V \Phi P_V
\end{equation}
where $P_V$ is the projector onto the valence subspace. 
Therefore the matrix $\S$ has an expansion of the form
\begin{equation}
\S = 1 + P_V \Phi P_V + \cdots
\end{equation}
Now, insert this expression into Eq.~\eqref{eq:LGL}, and consider
only the terms involving non-zero powers of $\Phi$. In the single
trace operators, the projectors remove any dependence on the sea
quark masses. There is still sea quark mass dependence remaining
in the double trace operators proportional to $L_4$ and $L_6$
given by
%\begin{widetext}
\begin{multline}
\delta \mathcal{L}_{GL} = 
	4 B_0\, L_4 \; \mathrm{str} \left( \partial_\mu \S P_V\, \partial^\mu \S^\dagger P_V \right) 
		\mathrm{str} (m_q) 
	\\ 
	+ 16 B_0^2\, L_6 \; \mathrm{str} \left(m_q \S^\dagger P_V + P_V\S m_q^\dagger\right)
		\mathrm{str} (m_q) .
\end{multline}
However, these operators simply shift $f$ and $B_0$
\begin{align}
f^2 & \rightarrow f^2 + 32 L_4\, B_0\, \mathrm{str} (m_q) \\
f^2 B_0 & \rightarrow f^2 B_0 + 64 L_6\, B_0^2\, \mathrm{str} (m_q).
\end{align}
Since the parameters $f$ and $B_0$ are eliminated in lattice-physical parameters in favor of the measured decay constants and meson masses, we can remove the dependence of scattering lengths on the sea quark masses by
working in lattice-physical parameters.  In an analogous way, we can remove all the explicit lattice spacing dependence.  The general MA Lagrangian involving valence-valence external states at $\mc{O}(\varepsilon_m^2 \varepsilon_a^2)$ can be reduced to the following form
%\begin{widetext}
\begin{align}
\delta \mc{L}_{MA} =&\ a^2 L_{a^2}^{\partial}\, 
	\str \left( \partial_\mu \S P_V\, \partial^\mu \S^\dagger P_V \right) 
	\nonumber\\ &\ 
	\times \str \left(  f(P_S \S P_S)\, f^\prime(P_S \S^\dagger P_S) \right) 
	\nonumber\\ & 
	+a^2 \, L_{a^2}^{m_q}\, \str \left(m_q P_V \S^\dagger P_V + P_V\S P_V m_q^\dagger\right) 
	\nonumber\\ &\
	\times \str \left( g(P_S \S P_S)\, g^\prime(P_S \S^\dagger P_S) \right)\, +h.c.,
\end{align}
%\end{widetext}
where the $f$'s and $g$'s are functions dependent upon the sea-quark lattice action.  These then lead to renormalizations of the LO constants,
\begin{align}
f^2 &\rightarrow f^2 
	+8a^2\, L_{a^2}^{\partial}\, \str \left( f(P_S \S P_S)\, f^\prime(P_S \S^\dagger P_S) \right)\, , 
	\nonumber\\
f^2 B_0 &\rightarrow 
	f^2 B_0 
	+4a^2\, L_{a^2}^{m_q}\, \str \left( g(P_S \S P_S) g^\prime(P_S \S^\dagger P_S) \right) ,
\end{align}
and just as with the sea quark mass dependence, expressing physical quantities in terms of the lattice-physical parameters removes any explicit dependence upon the lattice spacing in mesonic scattering processes.

Together, these results show that at NLO, the only counterterms
entering into the extrapolation formulae for mesonic scattering
lengths are the same as the counterterms entering into the physical
scattering length at NLO.  This lack of unphysical counterterms is
desirable from the point of view of chiral extrapolations, but it
also has another consequence. Loop graphs in quantum field theories
are frequently divergent; there must be a counterterm to absorb
these divergences in a consistent field theory. Since there is no
counterterm proportional to $a^2$ or the sea quark masses, loop
graphs involving these quantities are constrained so that they have no divergence proportional to $a^2$ or the sea quark masses.  This further reduces the possible sources of sea quark or lattice spacing
dependence. For example, mixed valence-sea meson masses have lattice
spacing shifts, so there can be no divergence involving the valence-sea
meson masses. In some cases this constraint is strong enough to
force the entire valence-sea mass dependence to cancel from scattering
lengths expressed in lattice-physical parameters. If this occurs,
then the scattering length will not depend on the unknown constant
$C_\mathrm{Mix}$. 

%%%%%%%%%%%%%%%%%%%%%%%%%%%%%%%%%%%%%%%%%%%%%%%%%%%
%
%	Finite Analytic Dependence
%
%%%%%%%%%%%%%%%%%%%%%%%%%%%%%%%%%%%%%%%%%%%%%%%%%%%
\subsection*{Dependence upon sea quarks \label{sec:MANLO_finite}}
Now, let us move on to discuss how the NLO extrapolation formulae
depend on the particular sea quark discretization in use. At NLO
in the effective field theory expansion, mesons composed of one
or two sea quarks only arise in loop graphs. In particular, the
valence-sea mesons can propagate between vertices where they
interact with valence-valence mesons; these interactions
involve the LO chiral Lagrangian augmented with the mixing term $a^2
\mathcal{U}_{VS}$. Because the mixing term is universal, these interaction
vertices are the same for all discretization schemes provided LO chiral
perturbation theory is applicable. The sea-sea mesons only arise
at NLO in hairpins; therefore, they are only sensitive to the quadratic
part of the appropriate LO chiral Lagrangian on the sea-sea
sector. Thus, we see that our NLO extrapolation formulae only depend
on the LO chiral Lagrangian to quadratic order in the sea-sea
sector and the LO chiral Lagrangian (with the mixing term) in the
valence-sea sector. Together, we see that the condition we
require on the sea quark discretization is that the sea-sea
sector alone should be described by chiral perturbation theory at LO,
and that the constant $C_\mathrm{Mix}$ should not be so large that its
explicit violation of chiral symmetry overwhelms the dynamical violation
of chiral symmetry. Non-locality which is described by the replica trick
does not present a problem since at the level of perturbation theory
the analytic continuation required by the replica method is trivial.
%
%However, at NLO these mesons simply propagate
%around a loop without interacting.  Therefore, the only way
%in which the sea quark discretization enters the NLO extrapolation
%formulae is through the lattice spacing dependent mass shifts of
%mesons containing the sea quarks, or through coefficients of
%unphysical operators appearing in the extrapolation formulae.
%

Note that the impact of using different sea quark discretizations
in our work is only at the level of the quadratic Lagrangian.
Therefore, the same NLO extrapolation formulae can be used to
describe simulations with different sea quark discretizations,
provided that the appropriate mass shifts are taken into account.  
In the case of staggered sea quarks, the sea-sea mass splitting which occurs in the MA formulae is that of the taste-identity, which has been measured~\cite{Aubin:2004fs}, and for the coarse MILC lattices, is given by
\begin{equation}
	a^2 \D_{sea} = a^2 \D_I \simeq (450 \textrm{ MeV})^2\, ,
\end{equation}
for $a \simeq 0.125$~fm.  These mass shifts can only appear through
the hairpin interactions at this order.  These terms will generally be
associated with \textit{unphysical} MA/PQ effects which give rise to
the enhanced chiral logarithms as well as additional finite analytic
dependence upon the sea-sea as well as valence-valence
meson masses (and their associated lattice spacing dependent mass
corrections).  The exception to this is the dependence upon the
$\eta$-mass.  As can be seen in Eq.~\eqref{eq:etaPropSU63}, the only way
the $\eta$-mass dynamically enters processes involving external pions and
kaons through $\mc{O}(\varepsilon_m^2 \varepsilon_a^2)$ is \textit{via}
the mass of the sea-sea $\eta$, Eq.~\eqref{eq:etaMass}.
The other way these discretization effects enter MA formulae is through
the mixed valence-sea meson masses, Eq.~\eqref{eq:LOMasses}.
Currently, this mass shift, $a^2 \D_\mathrm{Mix}$, is not known for any
type of sea quark discretization.  This is one of the more important
MA effects, because it enters many quantities of interest at the
one-loop level, for both mesons and baryons, and thus to perform chiral
extrapolations properly, this mass splitting must be taken into account.

%%%%%%%%%%%%%%%%%%%%%%%%%%%%%%%%%%%%%%%%%%%%%%%%%%%
%
%	MA NNLO: breakdown of continuum like behavior of Lagrangian
%
%%%%%%%%%%%%%%%%%%%%%%%%%%%%%%%%%%%%%%%%%%%%%%%%%%%
\subsection*{Mixed actions at NNLO}
It is important to note that these conclusions will not hold at NNLO in the effective field theory expansion. At this order, NNLO terms in the effective Lagrangian will introduce $a^2$ shifts of the Gasser-Leutwyler parameters themselves. In simulations which are precise enough to be sensitive to NNLO effects in chiral perturbation theory, these effects would have to be removed.  In addition, there will be new effects which can not be absorbed into the Gasser-Leutwyler parameters, but are truly new lattice spacing artifacts.  The simplest example to understand is to consider how the pion mass is modified at $\mc{O}(\varepsilon_m^4 \varepsilon_a^2)$ in a MA theory with staggered sea quarks~\cite{Walker-Loud:2006sa}.  

Briefly, there will be contributions to the pion mass which break taste, arising for example from the Gasser-Leutwyler operator in Eq.~\eqref{eq:LGL} with coefficient $L_6$.  The taste breaking contributions to the pion mass arise when the valence pion is contracted with the meson fields in one of the super-traces while the other super-trace is taken over sea-sea mesons which form a loop at this order,  
\begin{multline}
\d m_\pi^2(NNLO) = -\frac{64 m_\pi^2}{f^2}\, L_6 N_s^2
	\sum_{F,t} n_t\, \frac{B_0 (m_{s_1} +m_{s_2})}{(4\pi f)^2}	
	\\ 
	\times 
	\tilde{m}_{s_1 s_2,t}^2 \ln \left( \frac{\tilde{m}_{s_1 s_2,t}^2}{\mu^2} \right)\, ,
\end{multline}
where $N_s=1/4$ is the factor one inserts according to the replica method to account for the $4^{th}$-root of the sea quark determinant and $n_t$ counts the weighting of the mesons of various taste propagating in the loop.  The staggered meson mass of flavor $F$, and taste $t$, is given at LO by~\cite{Lee:1999zx,Aubin:2003mg,Aubin:2004fs}
\begin{equation}
\tilde{m}_{s_1 s_2,t}^2 = B_0 (m_{s_1} +m_{s_2}) +a^2 \Delta(\xi_t)\, .
\end{equation}
These taste-breaking effects are unphysical and their associated $\mu$-dependence can only be absorbed by the appropriate unphysical lattice spacing dependent operators arising in the mixed action Lagrangian.  This is simply one of many possible examples of how the continuum--like behavior of mixed action theories will break down.

%%%%%%%%%%%%%%%%%%%%%%%%%%%%%%%%%%%%%%%%%%%%%%%%%%%
%
%	Applications
%
%%%%%%%%%%%%%%%%%%%%%%%%%%%%%%%%%%%%%%%%%%%%%%%%%%%
\section{Applications}\label{sec:appl}

In this section, we discuss applications of these results to some specific quantities of physical interest.  
There have been a number of recent lattice computations~\cite{Bonnet:2004fr,Beane:2005rj,Edwards:2005ym,Beane:2006mx,Beane:2006pt,Beane:2006fk,Beane:2006kx,Beane:2006gj,Alexandrou:2006mc} utilizing the scheme first developed by the LHP collaboration~\cite{Renner:2004ck,Edwards:2005kw} of employing domain wall valence quarks with the publicly available MILC configurations.  In particular, the NPLQCD collaboration has computed the $I=2\ \pi\pi$ scattering length~\cite{Beane:2005rj},%
\footnote{In addition to the lattice spacing modifications of the $I=2\ \pi\pi$ scattering length computed in Ref.~\cite{Chen:2005ab}, the exponential finite volume corrections to this quantity were also computed in Ref.~\cite{Bedaque:2006yi}.  It was found that for the pion masses in use today, these effects were not significant, being on the order of 1\%.  It is expected that the exponential volume dependence in the other scattering processes will be similar to that of the two-pion system, as in all cases the pion is the lightest particle and will dominate the long range (finite volume) effects.} 
$f_K / f_\pi$~\cite{Beane:2006kx} and determined both the $I=3/2$ and $I=1/2\ K\pi$ scattering lengths through a direct determination of the $I=3/2\ K\pi$ scattering length~\cite{Beane:2006gj}.  As we will demonstrate by explicit computation, the $I=1\ KK$ scattering length, together with the above three systems share only two linearly independent sets of counterterms, which are the \textit{physical} counterterms of interest.  Therefore, these four quantities provide a means to test the mixed action formalism with only one lattice spacing.%
\footnote{In Ref.~\cite{Chen:2005ab}, it was argued that to all orders in perturbation theory, the unitarity violating features of MA and PQ theories do not invalidate the known method of extracting infinite volume scattering parameters from finite volume correlation functions~\cite{Hamber:1983vu,Luscher:1986pf,Luscher:1990ux}, for all ``maximally stretched" two-meson states, \textit{i.e.} the $I=2\ \pi\pi$, $I=3/2\ K\pi$ and $I=1\ KK$ systems.} 
%

%%%%%%%%%%%%%%%%%%%%%%%%%%%%%%%%%%%%%%%%%%%%%%%%%%%
%
%	App:  f_K / f_\pi
%
%%%%%%%%%%%%%%%%%%%%%%%%%%%%%%%%%%%%%%%%%%%%%%%%%%%
\subsection{$f_K / f_\pi$}

\begin{figure}
\center
\includegraphics[width=0.47\textwidth]{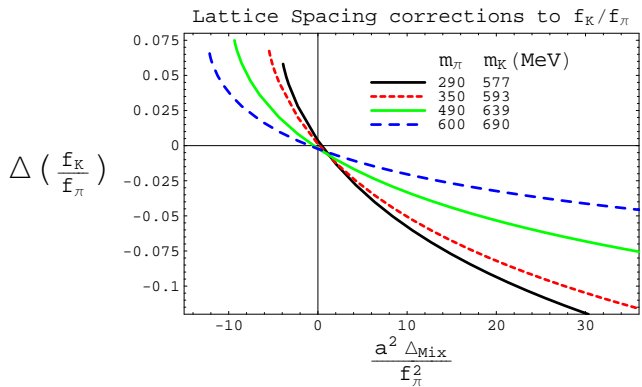}
\caption{\label{fig:fKfpiMAQCD}We plot the ratio, $\D(f_K / f_\pi)$ defined in Eq.~\eqref{eq:fKoverfpi} as a function of the unknown mixed meson mass splitting, $-(600 \textrm{ MeV})^2 \lesssim a^2 \D_\mathrm{Mix} \lesssim (800 \textrm{ MeV})^2$.  The observed deviation from the continuum $\chi$PT formulae is on the order of 5\%, which is important, but not significant enough to directly determine this unknown mass splitting from the MA lattice data of $f_K / f_\pi$~\cite{Beane:2006kx} alone.}
\end{figure}

The pion and kaon decay constants were computed in a mixed action theory with staggered sea quarks in Ref.~\cite{Bar:2005tu}.  In Appendix~\ref{app:decay_constants}, we include the general form of these results for arbitrary sea quarks to NLO, which we express in terms of the PQ parameters we introduced in Eq.~\eqref{eq:PQparameters}.  We use these formulae to estimate the error arising from the finite lattice spacing in the recent determination of $f_K / f_\pi$ in Ref.~\cite{Beane:2006kx}, in which the continuum $\chi$PT form of this quantity was used to extrapolate the lattice data to the physical point.  The MA functional form of this quantity depends upon the mixed valence-sea meson masses, and so we can not make a concrete prediction of the error made in this approximation, as the mixed meson mass depends upon $C_\mathrm{Mix}$, Eq.~\eqref{eq:LOMasses}, which is currently unknown.%
\footnote{In Ref.~\cite{Bunton:2006va}, this quantity was recently estimated by comparing the MA form of the pion form-factor to a MA simulation~\cite{Bonnet:2004fr}.  Unfortunately, only one of the lattice data points was in the chiral regime, so a precise determination of this quantity was not possible.} 
Consequently we form the ratio,
\begingroup
%\small
\begin{equation}\label{eq:fKoverfpi}
	\Delta \left( \frac{f_K}{f_\pi} \right) = 
	\frac{ \frac{f_K}{f_\pi} \bigg|_{MA} - \frac{f_K}{f_\pi} \bigg|_{QCD}}{\frac{f_K}{f_\pi} \bigg|_{QCD}}\, ,
\end{equation}
\endgroup
and in Eq.~\eqref{eq:fKfpiMAQCD}, we provide the explicit formula for this quantity, with the mass tuning used in Ref.~\cite{Beane:2006kx} ($m_{q_s} = m_{q_v} \Longrightarrow \tilde{\D}_{rs}^2 = \tilde{\D}_{ju}^2 = a^2 \D_I \simeq (450\ \textrm{MeV})^2$).  In Fig.~\ref{fig:fKfpiMAQCD}, we plot this ratio as a function of the mixed meson splitting in the range $-(600 \textrm{ MeV})^2 \lesssim a^2 \D_\mathrm{Mix} \lesssim (800 \textrm{ MeV})^2$.  We take the value of $L_5(\mu)$ from Ref.~\cite{Beane:2006kx}, as their various fitting procedures produced little variation in the extraction of $L_5(\mu)$.  This provides us with an indirect means at estimating the error in the extrapolation of the quantity $f_K / f_\pi$.  As can be seen from Fig.~\ref{fig:fKfpiMAQCD}, reasonable values of $a^2 \D_\mathrm{Mix}$ can produce deviations in $f_K / f_\pi$ on the order of 5\%.  These deviations are important enough to include in the fitting procedure (although still within the confidence levels in Ref.~\cite{Beane:2006kx,Beane:fKfpiPrivate}), but not significant enough to determine $a^2 \D_\mathrm{Mix}$ directly from the data in Ref.~\cite{Beane:2006kx}.  One can also determine the size of the hairpin contributions alone by setting $a^2 \D_\mathrm{Mix} =0$, and, as can be seen in Fig.~\ref{fig:fKfpiMAQCD}, these effects are a fraction of a percent for all values of the pion mass.

It is important to note that at this order, the counterterm structure of $f_K / f_\pi$ in a MA theory is identical to the counterterm structure of $f_K / f_\pi$ in $\chi$PT, as can be verified by examining Eqs.~\eqref{eq:fpi} and \eqref{eq:fK},
\begingroup
%\small
\begin{equation}
	\frac{f_K}{f_\pi} \bigg|_{MA} \propto \frac{8(m_K^2 -m_\pi^2)}{f^2}\, L_5(\mu)\, .
\end{equation}
\endgroup
This can be understood with the arguments presented in Sec.~\ref{sec:NLO}, and the knowledge that the lattice spacing artifacts are flavor-blind.

%%%%%%%%%%%%%%%%%%%%%%%%%%%%%%%%%%%%%%%%%%%%%%%%%%%
%
%	fig: fK over fpi - MA - QCD
%
%%%%%%%%%%%%%%%%%%%%%%%%%%%%%%%%%%%%%%%%%%%%%%%%%%%
%%%%%%%%%%%%%%%%%%%%%%%%%%%%%%%%%%%%%%%%%%%%%%%%%%%
%
%%%%%%%%%%%%%%%%%%%%%%%%%%%%%%%%%%%%%%%%%%%%%%%%%%%

%%%%%%%%%%%%%%%%%%%%%%%%%%%%%%%%%%%%%%%%%%%%%%%%%%%
%
%	App:  KK
%
%%%%%%%%%%%%%%%%%%%%%%%%%%%%%%%%%%%%%%%%%%%%%%%%%%%
\subsection{$KK\ I=1$ scattering length, $a_{KK}^{I=1}$}

Next, we discuss the form of the $I=1\ KK$ scattering length, for a MA theory with arbitrary sea quarks, for which the full functional form is provided in Appendix~\ref{app:kk}.  The two Kaon system is theoretically ideal for testing the convergence of $SU(3)$ $\chi$PT, however experimentally much more difficult to study.  But recent progress with lattice QCD simulations has allowed the $I=1\ KK$ system to be explored within the MA framework.  Thus one can use lattice QCD in combination with the appropriate MA effective field theory to explore the convergence of $SU(3)\ \chi$PT~\cite{Gasser:1984gg}, or whether a generalized version of $\chi$PT is a more appropriate description of nature~\cite{Stern:1993rg}.  In fact it has only recently been confirmed that the standard $SU(2)$ $\chi$PT power counting is phenomenologically correct~\cite{Colangelo:2000jc,Colangelo:2001sp,Colangelo:2001df}, by comparing our theoretical knowledge of the two-loop $\pi\pi$ scattering~\cite{Bijnens:1995yn,Bijnens:1997vq}, the pion scalar form-factor~\cite{Bijnens:1998fm} and the Roy equation analysis~\cite{Ananthanarayan:2000ht} with the recent experimental determination of the pion scattering lengths~\cite{Pislak:2001bf,Pislak:2003sv}.

The $I=1$ $KK$ system has several features in common with the $I=2$ $\pi\pi$ system discussed in Ref.~\cite{Chen:2005ab}. Firstly, the $I=1$ $KK$ system does not have on-shell hairpins in the $s$-channel loops.  Secondly, the scattering length does not depend upon the mixed valence-sea mesons when expressed in terms of the lattice-physical parameters, and finally, the only counterterm at NLO is the physical counterterm of interest. The form of the $I=1\ KK$ scattering length is given by
%\begin{widetext}
\begin{multline}\label{eq:aKK}
m_K a_{KK}^{I=1} =-\frac{m_{K}^{2}}{8 \pi f_{K}^{2}} \bigg\{ 1
	+\frac{m_{K}^{2}}{ (4\pi f_{K})^{2}} \bigg[ 
		C_\pi \ln \left( \frac{m_\pi^{2}}{\mu^2} \right) \\
		+C_K \ln \left( \frac{m_K^{2}\,}{\mu ^2} \right) 
		+C_X \ln \left( \frac{\tilde{m}_X^{2}}{\mu^2} \right) 
		+C_{ss} \ln \left( \frac{m_{ss}^{2}}{\mu ^{2}} \right) \\
		+ C_0 
	- 32(4\pi)^2\, L_{KK}^{I=1}(\mu)
	\bigg] \bigg\}\, ,
\end{multline}
%\end{widetext}
where the various coefficients, $C_\phi$, are provided in Eqs.~\eqref{eq:KKCpi}--\eqref{eq:KKC0}.

One important point is that the counterterm for the $I=1$ scattering length, $L_{KK}^{I=1}$ is identical to the $I=2\ \pi\pi$ scattering length counterterm,
\begingroup
%\small
\begin{align}\label{eq:LKK}
L_{KK}^{I=1} &= L_{\pi\pi}^{I=2} \nonumber\\
	&= 2L_1 +2L_2 +L_3 -2L_4 -L_5 +2L_6 +L_8\, .
\end{align}
\endgroup
Before discussing this scattering length in more detail, we first give the result in $\chi$PT, as this has not been presented in the literature to the authors' knowledge.  
%\begin{widetext}
%\small
\begin{align}
m_K a_{KK}^{I=1} =&\ -\frac{m_K^2}{8 \pi f_K^2} \bigg\{ 1 +\frac{m_K^2}{(4 \pi f_K)^2} \bigg[
	2 \ln \left( \frac{m_K^2}{\mu^2} \right)
	\nonumber\\ &\ 
	-\frac{2 m_\pi^2}{3 (m_\eta^2 -m_\pi^2)}\, \ln \left( \frac{m_\pi^2}{\mu^2} \right)
	\nonumber\\ &\ 
	+\frac{2(20m_K^2 -11m_\pi^2)}{27(m_\eta^2 -m_\pi^2)}\,
	\ln \left( \frac{m_\eta^2}{\mu^2} \right)
	\nonumber\\ &\ 
	-\frac{14}{9}
	-32(4\pi)^2\, L_{KK}^{I=1}(\mu)
	\bigg]
	\bigg\}\, ,
\end{align}
%\end{widetext}
with $L_{KK}^{I=1}$ given in Eq.~\eqref{eq:LKK}, and we have used the leading order meson mass relations to simplify the form of this expression.

The equality of the $I=2\ \pi\pi$ and $I=1\ KK$ scattering length counterterms allows us to make a prediction for the numerical values of $m_K a_{KK}^{I=1}$ one should obtain in a simulation of this system with domain-wall valence quarks on the MILC configurations.  To do this, we must first convert the counterterm, $l_{\pi\pi}(\mu)$ obtained by NPLQCD in Ref.~\cite{Beane:2005rj} from the  effective theory with two sea flavors to the theory with three sea flavors.  For PQ and MA theories, there is an additional subtlety which arises in this matching.  If we match the $\chi$PT forms of $m_\pi a_{\pi\pi}^{I=2}$ in $SU(2)$ to $SU(3)$, then we arrive at the equality (with the conventions defined in Appendix~\ref{app:pipi})
\begingroup
%\small
\begin{equation}\label{eq:SU2SU3matching}
l_{\pi\pi}^{I=2}(\mu) = 32 (4\pi)^2\, L_{\pi\pi}^{I=2}(\mu) - \frac{1}{9}\, \ln \left( \frac{m_\eta^2}{\mu^2} \right)
	-\frac{1}{9}\, .
\end{equation}
\endgroup
This leads to an exact matching between the $SU(2)$ and $SU(3)$ theories, in which all of the strange quark mass dependence at this order, which is purely logarithmic, is absorbed in the $SU(2)$ Gasser-Leutwyler coefficients~\cite{Gasser:1984gg}.  %
%%%%%%%%%%%%%%%%%%%%%%%%%
%
%	Table: I=2 pipi PQ Effects
%
%%%%%%%%%%%%%%%%%%%%%%%%%
\begin{table*}
\caption{\label{t:Fn} Hairpin contributions to $m_\pi a_{\pi\pi}^{I=2}$.  We provide the various hairpin contributions to the $I=2\ \pi\pi$ scattering length for both the 2-sea flavor, $(b)$, and 3-sea flavor theory, $(d)$, which we compare to the $\chi$PT NLO contribution, $(a)$ and LO contribution, top-row.  In row $(c)$, we give the new hairpin effects which arise in the 3-flavor theory, and in $(d)$ we provide the total 3-sea flavor hairpin effects.}
\begin{tabular}{| c | c | c c c c |}
\hline
&  $m_\pi$ (MeV) & $293$ & $354$ & $493$ & $592$  \\
 \hline
& $-\frac{m_\pi^2}{8\pi f_\pi^2}$ & $-0.156$ & $-0.218$ & $-0.372$ & $-0.483$ \\
$(a)$ & $ -\frac{2\pi m_\pi^4}{(4\pi f_\pi)^4} \Big[ 3 \ln \left( \frac{m_\pi^2}{\mu^2} \right) -1 -l_{\pi\pi}^{I=2}(\mu) \Big]$ 
	& $0.00460$ & $0.00140$ & $-0.0314$ & $-0.0818$ \\
$(b)$ & $-\frac{2\pi m_\pi^4}{(4\pi f_\pi)^4} \Big[ -\frac{\tilde{\D}_{ju}^4}{6 m_\pi^4} \Big]$ & $0.00359$ & $0.00327$ & $0.00254$ & $0.00207$ \\
$(c)$ & $ -\frac{2\pi m_\pi^4}{(4\pi f_\pi)^4} \Big[ \sum_{n=1}^4 \left( \frac{\tilde{\D}_{ju}}{m_\pi^2} \right)^n \mc{F}_n (m_\pi^2 / \tilde{m}_X^2) \Big]$ & $-0.00243$ & $-0.00289$ & $-0.00371$ & $-0.00396$ \\
$(d)$ & $-\frac{2\pi m_\pi^4}{(4\pi f_\pi)^4} \Big[ -\frac{\tilde{\D}_{ju}^4}{6 m_\pi^4} +
 \sum_{n=1}^4 \left( \frac{\tilde{\D}_{ju}}{m_\pi^2} \right)^n \mc{F}_n (m_\pi^2 / \tilde{m}_X^2) \Big]$  
	& $0.00116$ & $0.00040$ & $-0.00117$ & $-0.00188$ \\
\hline
\end{tabular}
%\end{ruledtabular}
\end{table*}
If we naively attempt to match the $SU(4|2)$ to $SU(6|3)$ MA/PQ expressions for $m_\pi a_{\pi\pi}^{I=2}$, using Eqs.~\eqref{eq:I2pipiSU2} and \eqref{eq:I2pipiSU3}, one arrives at the relation
\begin{align}\label{eq:PQmatching}
l_{\pi\pi}^{I=2}(\mu) =&\ 32 (4\pi)^2\, L_{\pi\pi}^{I=2}(\mu) - \frac{1}{9}\, 
	\ln \left( \frac{\tilde{m}_X^2}{\mu^2} \right)
	-\frac{1}{9} 
	\nonumber\\
	&\ -\sum_{n=1}^4\, 
	\left( \frac{\tilde{\D}_{ju}^2}{m_\pi^2} \right)^n\, \mc{F}_n (m_\pi^2 / \tilde{m}_X^2)\, ,
\end{align}
where the functions, $\mc{F}_n(y)$ were first determined in Ref.~\cite{Chen:2005ab} and are given in Eqs.~\eqref{eq:coolFs}. All of the new terms in this matching arise from the extra hairpin interactions present in the $SU(6|3)$ theory which are not present in $SU(4|2)$.  One can show that these terms are formally higher order in the $SU(4|2)$ chiral expansion, but nevertheless we will see that they are not negligible.

The NPLQCD collaboration has recently computed $m_\pi a_{\pi\pi}^{I=2}$ and used the $SU(2)$ extrapolation formula to determine $l_{\pi\pi}^{I=2}$~\cite{Beane:2005rj}.  Adjusting for conventions and including their largest uncertainty, they determined
\begin{equation}\label{eq:lpipiNPLQCD}
	l_{\pi\pi}^{I=2}(4 \pi f_\pi) \simeq -10.9 \pm1.8\, .
\end{equation}
Starting with this determination, we can then compare the hairpin contributions in $SU(4|2)$ to those of $SU(6|3)$ and also to the \textit{physical} contribution at NLO.
%, being the combination of the counterterm and the logarithm in Eq.~\eqref{eq:2seaScattLength}, which we collect in Table~\ref{t:Fn}.  
We collect these results in Table~\ref{t:Fn}.

%What we find is very interesting.  
The two-flavor hairpin effects, listed in row $(b)$ of Table~\ref{t:Fn}, are not small relative to the (scale independent) $\chi$PT NLO contributions $(a)$, and for the lightest two masses shown, are of the same order of magnitude.  However, when we consider the three flavor theory, we see that the additional hairpin effects, $(c)$, are of the same order as the two flavor hairpin contributions (which also contribute in the three flavor theory), but opposite in sign. 
Taking into account all of the hairpin contributions by using the three flavor theory, one observes that the sum of these unphysical effects, $(d)$, is approximately an order of magnitude smaller than the physical NLO effects of $\chi$PT, but increases in importance as the pion mass is reduced. This justifies our assumption of the determination of $l_{\pi\pi}^{I=2}$ given in Eq.~\eqref{eq:lpipiNPLQCD}. This also justifies the $SU(3) \rightarrow SU(2)$ matching given in Eq.~\eqref{eq:SU2SU3matching}, and explains the success found in Ref.~\cite{Beane:2005rj} of using the $SU(2)\ \chi$PT formula to determine $l_{\pi\pi}^{I=2}$, as the unphysical hairpin corrections to this formula provide a relative shift of about 10\% to the NLO contributions for the masses simulated, which is roughly the size of their largest quoted error.  

\begin{figure}
\includegraphics[width=0.49\textwidth]{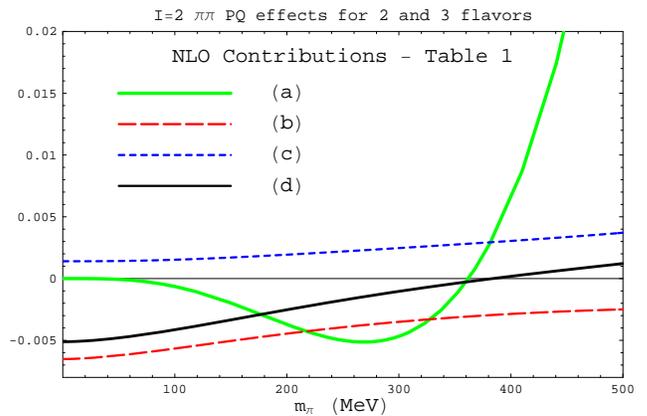}
\caption{\label{fig:PQpipi} We plot the absolute values of the various NLO contributions to $m_\pi a_{\pi\pi}^{I=2}$ listed in Table~\ref{t:Fn}.  The NLO $\chi$PT contribution is given by $(a)$ (green), which demonstrates the large cancellation of the counterterm and the chiral log for light to medium pion masses.  The long-dashed curve (red) is the 2-sea flavor hairpin effects, $(b)$, which are the same order of magnitude as $(a)$, for $m_\pi \lesssim 400\ \textrm{MeV}$.  When the new 3-sea flavor hairpin effects, $(c)$ (blue), are added to the 2-sea flavor effects, one finds that the total 3-sea flavor hairpin effects, $(d)$ (black), are small compared to $(a)$ for $m_\pi \gtrsim 250$~MeV.}
\end{figure}

In Fig.~\ref{fig:PQpipi}, we plot the absolute values of the various NLO contributions to the $I=2\ \pi\pi$ scattering length as a function of the pion mass, which highlight the importance of these hairpin effects.  Their relative importance is enhanced for the values listed in Table~\ref{t:Fn} because of the large cancellation of the counterterm and chiral log at NLO, $(a)$.  It is clear that these effects will become more important as one moves further into the chiral regime, $(m_\pi \rightarrow 0)$.
%%%%%%%%%%%%%%%%%%%%%%%%%
%
%	Figure:PQ Effects
%
%%%%%%%%%%%%%%%%%%%%%%%%%
%
%
%

%%%%%%%%%%%%%%%%%%%%%%%%%
%
%	Table: I=1 KK PQ Effects and prediction
%
%%%%%%%%%%%%%%%%%%%%%%%%%
\begin{table*}[t]
\caption{\label{t:mKaKK} Predictions of $m_K a_{KK}^{I=1}$.  We use the equality of the $m_K a_{KK}^{I=1}$ and $m_\pi a_{\pi\pi}^{I=2}$ counterterms (expressed in lattice-physical parameters) to predict the values of $m_K a_{KK}^{I=1}$ which would be computed in a MA lattice formulation with domain wall valence quarks on the MILC staggered sea quarks, which we compare both to the tree level prediction as well as the $SU(3)\ \chi$PT prediction, for values of $m_K / f_K$ taken from Refs.~\cite{Beane:2005rj,Beane:2006kx}.  We also provide a prediction of the scattering length at the physical point.  
The first error is due to the uncertainty in the determination of $L_{KK}^{I=1}$ from Eq.~\eqref{eq:PQmatching} and the value of $l_{\pi\pi}^{I=2}$, Eq.~\eqref{eq:lpipiNPLQCD} determined in Ref.~\cite{Beane:2005rj}.  The second error is a power counting estimate of the NNLO contributions to the scattering length.}
%\begin{ruledtabular}
\begin{tabular}{| c | c c c c |}
\hline
$m_K: f_K$ (MeV) & $577 : 172$ & $593 : 171$ & $639 : 173$ & $690 : 177$ \\
 \hline
$m_K a_{KK}^{I=1}$(LO): $-\frac{m_K^2}{8\pi f_K^2}$ & $-0.447$ & $-0.479$ & $-0.542$ & $-0.605$ \\
$m_K a_{KK}^{I=1}$(NLO: MA) & $-0.091$ & $-0.113$ & $-0.162$ & $-0.223$ \\
$m_K a_{KK}^{I=1}$(NLO: $SU(3)$) & $-0.084$ & $-0.107$ & $-0.157$ & $-0.217$ \\
\hline
$m_K a_{KK}^{I=1}$(MA) & $-0.540 \pm0.069 \pm 0.026$ & $-0.592 \pm 0.079 \pm 0.031$ & 
	$-0.704 \pm 0.102 \pm 0.048$ & $-0.828 \pm 0.127 \pm 0.072$ \\
$m_K a_{KK}^{I=1}$($SU(3)$) & $-0.531 \pm0.069 \pm 0.026$ & $-0.586 \pm 0.079 \pm 0.031$ 
	& $-0.699 \pm 0.102 \pm 0.048$ & $-0.823 \pm 0.127 \pm 0.072$
	 \\
\hline
\hline
physical point & $496 : 161$ &&& \\
\hline
$m_K a_{KK}^{I=1} (SU(3))$ & $-0.424 \pm 0.049 \pm 0.012$ &&& \\
\hline
\end{tabular}
%\end{ruledtabular}
\end{table*}
Given the small contribution of the NLO hairpin effects to $m_\pi a_{\pi\pi}^{I=2}$, we can use the determination of $l_{\pi\pi}^{I=2}(\mu)$ in Eq.~\eqref{eq:lpipiNPLQCD}~\cite{Beane:2005rj}, and the matching of Eq.~\eqref{eq:SU2SU3matching} to determine $L_{KK}^{I=1}(\mu)$ and thus predict values of $m_K a_{KK}^{I=1}$, which we provide in Table~\ref{t:mKaKK}. We provide both the comparison of the NLO effects as predicted by both the MA theory as well as $SU(3)\ \chi$PT, which we 
compare to the tree level prediction, as well as the total scattering length through NLO.  We find that 
similar to $m_\pi a_{\pi\pi}^{I=2}$, the NLO hairpin effects for $m_K a_{KK}^{I=1}$ are only about 10\% of 
the NLO $\chi$PT value, less than the accuracy we claim here.  We find that a current MA lattice determination of $m_K a_{KK}^{I=1}$ will not be sensitive to the \textit{unphysical} hairpin contributions with the expected level of uncertainty, as can be seen by the predicted MA and $SU(3)$ values.  However, for both the MA and $SU(3)$ theories, the NLO contributions are 15--30\% correction to the LO term showing a convergence expected by power counting.

The first error is due to the uncertainty in the determination of $L_{KK}^{I=1}$ we obtain from the matching in Eq.~\eqref{eq:PQmatching} and the extraction of $l_{\pi\pi}^{I=2}$ from Ref.~\cite{Beane:2005rj}, Eq.~\eqref{eq:lpipiNPLQCD}.  This uncertainty includes estimations of the two-loop contributions to $m_\pi a_{\pi\pi}^{I=2}$ in $SU(2)$ $\chi$PT.  The second uncertainty listed in Table~\ref{t:mKaKK} is a power counting estimation of the NNLO contributions to $m_K a_{KK}^{I=1}$.  Some of these effects are already included in the first uncertainty but a conservative estimate of our predicted error is to add these uncertainties in quadrature.

%%%%%%%%%%%%%%%%%%%%%%%%%%%%%%%%%%%%%%%%%%%%%%%%%%%
%
%	App:  Kpi
%
%%%%%%%%%%%%%%%%%%%%%%%%%%%%%%%%%%%%%%%%%%%%%%%%%%%
\subsection{$K\pi\ I=3/2$ scattering length, $a_{K\pi}^{I=3/2}$}

The $K\pi$ system is also an interesting laboratory for exploring the three flavor structure of low energy hadron interactions, and moreover it is experimentally accessible with proposed studies by the DIRAC collaboration~\cite{DIRAC}.  There has recently been a direct MA lattice QCD determination of the $I=3/2\ K\pi$ scattering length, which in combination with the theoretical knowledge of the NLO $\chi$PT  $I=1/2$ and $I=3/2\ K\pi$ scattering lengths~\cite{Bernard:1990kw,Bernard:1990kx,Kubis:2001ij,Kubis:2001bx} has allowed a determination of both isospin scattering lengths~\cite{Beane:2006gj}.  There is additionally a two-loop computation of $K\pi$ scattering in $SU(3)$ $\chi$PT which studies the convergence of the theory with standard power counting~\cite{Bijnens:2004bu}.  Before embarking on a study of the two-loop effects with lattice QCD, one must first understand the lattice corrections at NLO.  This is the motivation for this section.

The tree level $I=3/2\ K\pi$ scattering length is given by
\begin{align}
	(m_\pi+m_K) a_{K\pi}^{I=3/2} &= -\frac{m_\pi m_K}{4\pi f_K f_\pi}\, ,\textrm{ or} \nonumber\\
	\mu_{K\pi} a_{K\pi}^{I=3/2} &= -\frac{\mu_{K\pi}^2}{4\pi f_K f_\pi}\, ,
\end{align}
where $\mu_{K\pi}$ is the reduced mass of the $K\pi$ system.  We chose to express our extrapolation
formulae in terms of the product $f_K f_\pi$ since this symmetric treatment
of the $K$ and $\pi$ mesons provides the simplest form of the scattering length.  We find, however, that the $I=\frac{3}{2}$ scattering length still depends on the mixed valence-sea meson masses, and therefore on the parameter $C_\mathrm{Mix}$.
Consequently, accurate chiral extrapolations of this scattering
length will require a determination of the value of $C_\mathrm{Mix}$ appropriate
to the particular sea quark discretization used in the simulation.  The form of the MA $I=3/2\ K\pi$ scattering length is
\begin{align}
\mu_{K\pi} a_{K\pi}^{I=3/2} =&\ -\frac{\mu_{K\pi}^2}{4\pi f_K f_\pi}\, \bigg[
	1 -\frac{32 m_Km_\pi}{f_K f_\pi}\, L_{\pi\pi}^{I=2}(\mu)
	\nonumber\\ &\ 
	+\frac{8(m_K -m_\pi)^2}{f_K f_\pi}\, L_5(\mu) \bigg]
	\nonumber\\ &\ 
	+\mu_{K\pi} \Big[ a_{vv}^{K\pi,3/2}(\mu) +a_{vs}^{K\pi,3/2}(\mu) \Big]\, ,
\end{align}
where $a_{vv}^{K\pi,3/2}(\mu)$ is the valence-valence (including valence-ghost) contribution to the scattering length and $a_{vs}^{K\pi,3/2}(\mu)$ is a non-vanishing contribution from mixed valence-sea mesons to the scattering length.  The other important thing to note is that there are two counterterms for this scattering length which can both be determined through its chiral extrapolation formula, but can also independently be determined in other processes; $L_5(\mu)$ can be determined independently by $f_K / f_\pi$ and $L_{\pi\pi}^{I=3/2}(\mu)$ can be determined either with $I=2\ \pi\pi$ or $I=1\ KK$ scattering.  Now we have explicitly demonstrated that the four observable quantities, $a_{\pi\pi}^{I=2}$, $a_{KK}^{I=1}$, $a_{K\pi}^{I=3/2}$ and $f_K / f_\pi$, when expressed in terms of the lattice-physical parameters, only share two linearly independent counterterms through NLO in MA (and PQ) $\chi$PT.

Before continuing, we provide the continuum $SU(3)$ $\chi$PT form of $a_{K\pi}^{I=3/2}$, which is the same as can be constructed from Ref.~\cite{Kubis:2001bx} with the NLO shift of $f_K \rightarrow f_\pi$,
\begin{widetext}
\begin{multline}
\mu_{K\pi}\, a_{K\pi}^{I=3/2} =\ -\frac{\mu_{K\pi}^2}{4\pi f_K f_\pi}\, \Bigg\{
	1 + \frac{1}{(4\pi)^2 f_K f_\pi}\, \Bigg[
		\kappa_\pi\, \ln \left( \frac{m_\pi^2}{\mu^2} \right)
%	\\
		+\kappa_K\, \ln \left( \frac{m_K^2}{\mu^2} \right)
		+\kappa_\eta\, \ln \left( \frac{m_\eta^2}{\mu^2} \right)
		-\frac{86}{9}\, m_Km_\pi 
	\\
%	\nonumber\\ &\qquad\qquad
		+\kappa_{tan}\, \arctan \left( \frac{2(m_K -m_\pi) \sqrt{2 m_K^2 +m_K m_\pi -m_\pi^2}}
			{(2m_K -m_\pi)(m_K +2m_\pi)} \right) \Bigg] 
	\\
%	\nonumber\\ &\qquad
		-\frac{32 m_K m_\pi}{f_K f_\pi}\, L_{\pi\pi}^{I=2}(\mu)
		+\frac{8(m_K -m_\pi)^2}{f_K f_\pi}\, L_5(\mu) 
	\Bigg\}\, ,
\end{multline}
\end{widetext}
with
\begin{align}
	\kappa_\pi &=  -\frac{m_\pi^2}{4}\frac{11m_K^2 +22m_Km_\pi -5m_\pi^2}{m_K^2 -m_\pi^2}\, , \\
	\kappa_K &= \frac{m_K}{18}\frac{ 134m_K^2 m_\pi -9m_K^3 +55m_Km_\pi^2-16m_\pi^3}
		{m_K^2 -m_\pi^2}\, , \\
	\kappa_\eta &= \frac{-36m_K^3 -12m_K^2m_\pi +m_Km_\pi^2 +9m_\pi^3}{36(m_K-m_\pi)}\, , \\
	\kappa_{tan} &= \frac{16 m_Km_\pi}{9}\frac{\sqrt{2m_K^2 +m_Km_\pi -m_\pi^2}}{m_K-m_\pi}\, .
\end{align}

In Appendix~\ref{app:kpi}, we provide the full form of the MA $I=3/2\ K\pi$ scattering length.  Here we wish to examine the valence-sea contribution in more detail.  This contribution to the scattering length is given by%
\footnote{We note that the summation over sea flavor, $F$, implicitly includes the appropriate factors for staggered sea quarks, the sum over taste and the factors of $N_s=1/4$ which arise from the $4^{th}$-rooting trick.  For other brands of quarks, this is simply a sum over the sea flavors, $j, l$ and $r$.} 
\begin{multline}
\label{eq:aKpiVS}
\mu_{K\pi}\, a_{vs}^{K\pi,3/2}(\mu) = -\frac{\mu_{K\pi}^2}{4\pi f_K f_\pi} \frac{1}{2(4\pi)^2 f_K f_\pi}
	\sum_{F=j,l,r} 
	\\ \times 
	\bigg[ C_{Fs} \ln \left( \frac{\tilde{m}_{Fs}^2}{\mu^2} \right)
		-C_{Fd} \ln \left( \frac{\tilde{m}_{Fd}^2}{\mu^2} \right)
	\\
	+4m_K m_\pi J(\tilde{m}_{Fd}^2)
		\bigg]\, ,
\end{multline}
where
\begin{align}
	C_{Fs} &= \frac{4 m_K^2 m_\pi - \tilde m_{Fs}^2 ( m_K + m_\pi)}{m_K - m_\pi}\, , \label{eq:CFs}\\
	C_{Fd} &= \frac{4 m_K m_\pi^2 - \tilde m_{Fd}^2 ( m_K + m_\pi)}{m_K - m_\pi}\, , \label{eq:CFd}\\
	J(M) &= 2 \frac{\sqrt{M^2 - m_\pi^2}}{m_K - m_\pi}\, 
		\arctan \left( \frac{(m_K - m_\pi) \sqrt{M^2 - m_\pi^2}}{M^2 + m_K m_\pi -m_\pi^2} \right)
	\nonumber\\ &\qquad
		-m_K m_\pi\, .\label{eq:JVS}
\end{align}
The scale dependence in $a_{vs}^{K\pi,3/2}(\mu)$ can be shown to be
\begin{equation}\label{eq:a32VSmu}
a_{vs}^{K\pi,3/2}(\mu) \propto \sum_{F=j,l,r} -\ln (\mu^2) (m_K -m_\pi)^2\, ,
\end{equation}
and as claimed, independent of both the lattice spacing, $a$, and the sea quark masses, and is  absorbed by $L_5(\mu)$.  We stress again that the $I=3/2\ K\pi$ scattering length depends upon mixed valence-sea mesons, which receive lattice spacing dependent mass shifts proportional to the unknown quantity, $C_\mathrm{Mix}$.  This quantity is currently unknown for all variants of MA lattice QCD and must be determined for a correct extrapolation of MA lattice QCD simulations.  For this reason, we do not provide a table with \textit{post}-dictions of $\mu_{K\pi} a_{K\pi}^{I=3/2}$.

%%%%%%%%%%%%%%%%%%%%%%%%%%%%%%%%%%%%%%%%%%%%%%%%%%%
%
%	Discussion and suggestions
%
%%%%%%%%%%%%%%%%%%%%%%%%%%%%%%%%%%%%%%%%%%%%%%%%%%%
\section{Discussion}\label{sec:concl}

Mixed action simulations provide a promising solution to the problem
of performing fully dynamical simulations with light quarks which are
under theoretical control. Recently, several simulations have been
performed using the publicly available MILC staggered lattices with
domain-wall valence quarks and the future for simulations using Wilson
sea quarks and GW valence quarks seems to be bright.  This work shows
that mixed action simulations with Ginsparg-Wilson valence quarks are
theoretically clean. We have shown that the counterterms appearing in
mesonic scattering lengths are precisely those that occur in QCD, so that
one can in principle measure these counterterms with a single lattice
spacing. We also find that the same chiral extrapolation formulae can be
used to describe mixed action simulations with GW quarks with mild
restrictions on the type of sea quark discretization used---provided,
of course, that QCD is recovered in the continuum limit. Thus, our
results hold for simulations with domain wall and overlap quarks, Wilson quarks ($\mc{O}(a)$ improved and twisted mass quarks at maximal twist) as well as simulations using rooted staggered
quarks (assuming that the $4^{th}$-rooting procedure is valid and that the
replica method correctly captures all of the non-locality introduced by
the rooting procedure - which has been argued to all orders in perturbation
theory~\cite{Bernard:2006zw}).

It was previously observed in Ref.~\cite{Chen:2005ab} that the $\pi\pi$ scattering
length does not depend on the parameter $C_\mathrm{Mix}$ of mixed
action chiral perturbation theory. Here, we find that this also
holds for the $KK$ scattering length. However, the $K\pi$ scattering
length does depend on $C_\mathrm{Mix}$, 
and, therefore, accurate chiral extrapolations of mixed action data will require a measurement of
this quantity.  However, we have also computed the ratio $f_K/f_\pi$ in mixed action
chiral perturbation theory, which depends upon on $C_\mathrm{Mix}$.  By varying $C_\mathrm{Mix}$ over a broad range of values, we find the impact to be modest, on the order of 5\%. In addition, taking into account the small hairpin corrections to $m_\pi a_{\pi\pi}^{I=2}$~\cite{Chen:2005ab} and the equally small predicted corrections to $m_K a_{KK}^{I=1}$, Table~\ref{t:mKaKK}, we expect the impact of $C_\mathrm{Mix}$ on $a_{K\pi}^{I=3/2}$ to also be small at this order of precision, $\mc{O}(\varepsilon_m^4, \varepsilon_m^2 \varepsilon_a^2, \varepsilon_a^4)$.  In Table~\ref{t:mKaKK} we provide predictions of $m_K a_{KK}^{I=1}$ for various values of $m_K$ and also provide a prediction at the physical point.  

In Sec.~\ref{sec:NLO}, we have demonstrated why the use of lattice-physical parameters (or on-shell renormalization) significantly simplifies the form of the extrapolation formulae for mesonic systems.  We stress that these arguments do not depend upon the momentum of the system, nor upon having only two external mesons, and thus will be applicable not just for scattering lengths, but also for other scattering parameters, such as the effective range, as well as for $N > 2$ mesonic systems.  In the appendices we have provided explicit NLO extrapolation formulae for the meson masses and decay constants as well as the three scattering lengths discussed in this paper, for arbitrary sea quark discretization schemes, expressed in terms of the PQ parameters we introduced in Eq.~\eqref{eq:PQparameters}.  A thorough understanding of the lattice spacing effects at this order will require knowledge of the counterterms in the masses and decay constants.

We would like to conclude with a small point and a few suggestions.  If one is interested in removing the unitarity violating effects in MA lattice simulations, for the low-energy dynamics of the system, then theoretical analysis unambiguously advocates the tuning $\tilde{\D}_{rs} = \tilde{\D}_{ju} = 0$, which is the generalization of $m_{q_{sea}} - m_{q_{val}} = 0$ for PQ theories.  This is the most QCD-like scenario for MA theories in which the unitarity violating double pole propagators in Eq.~\eqref{eq:etaPropSU63} are tuned to zero.  It has recently been shown that this double pole structure of the flavor neutral propagators persists to all orders in PQ$\chi$PT~\cite{Bijnens:2006ca}, and thus this will be the appropriate tuning to higher orders as well.  From the point of view of doing chiral physics, this is not desirable for the coarse MILC lattices, as the lattice spacing shift to the \textit{taste-identity} staggered mesons is $a^2 \D_I \simeq (450 \textrm{ MeV})^2$, which would make for heavy pions.  Therefore we caution users of MA lattice simulations to remember the existence of these unitarity violating effects present in current MA simulations.

The simplified form of MA/PQ extrapolation formulae for the two-meson systems is particularly dependent upon the implications of the chiral symmetry of the valence quarks.  However, we conjecture that a similar, but not as strong, simplification will occur for other hadronic observables, in particular for nuclear physics as well as heavy meson observables, if also expressed in terms of lattice-physical parameters, which will lead to improved chiral extrapolations.  This is supported by the recent fits of the NPLQCD collaboration~\cite{Beane:2005rj,Beane:2006mx,Beane:2006pt,Beane:2006fk,Beane:2006kx,Beane:2006gj} and the LHP collaboration~\cite{Edwards:2006qx}.
Based upon our theoretical understanding of effective field theories designed to incorporate lattice spacing artifacts, we expect that even for fermion discretization schemes which do not have chiral symmetry, the use of lattice-physical parameters (on-shell renormalization) will in general simplify the chiral extrapolation formulae and improve chiral fits.

%%%%%%%%%%%%%%%%%%%%%%%%%%%%%%%%%%%%%%%%%%%%%%%%%%%
%
%	Acknowledgments
%
%%%%%%%%%%%%%%%%%%%%%%%%%%%%%%%%%%%%%%%%%%%%%%%%%%%
\begin{acknowledgments}
We would like to thank Martin Savage for many useful discussions. We also
thank Ruth Van de Water for her input at the start of this project. JWC is
supported by the National Science Council of R.O.C.. DOC is supported in
part by the U.S. DOE under the grant DE-FG03-9ER40701.  AWL was supported
under DOE grants DE-FG03-97ER-41014 and DE-FG02-93ER-40762.

\end{acknowledgments}

%%%%%%%%%%%%%%%%%%%%%%%%%%%%%%%%%%%%%%%%%%%%%%%%%%%
%
%	Appendices
%
%%%%%%%%%%%%%%%%%%%%%%%%%%%%%%%%%%%%%%%%%%%%%%%%%%%
\appendix

%%%%%%%%%%%%%%%%%%%%%%%%%%%%%%%%%%%%%%%%%%%%%%%%%%%
%
%		Appendix: pionic quantities in 2-flavors
%
%%%%%%%%%%%%%%%%%%%%%%%%%%%%%%%%%%%%%%%%%%%%%%%%%%%
\section{$m_\pi$ and $f_\pi$ for 2 sea flavors \label{app:2flav_mpi_fpi}}
In this Appendix, we provide the explicit formulae for the pion mass and decay constant in a two-sea flavor MA theory.  These were first computed in Refs.~\cite{Bar:2003mh,Bar:2005tu}.  Here we provide the answers expressed in terms of the PQ parameters we introduced in Eq.~\eqref{eq:PQparameters}.

\begin{widetext}
\begin{align}
m_\pi^2 =&\ 2B_0 \hat{m} \bigg\{ 1 + \frac{m_\pi^2}{(4\pi f)^2} \ln \left( \frac{m_\pi^2}{\mu^2} \right)
		-\frac{m_\pi^2}{f^2} \ell^{(m)}(\mu) 
%\nonumber\\ &\qquad\qquad
		-\frac{\tilde{\D}_{ju}^2}{(4\pi f)^2}\, \left[ 1+ \ln \left( \frac{m_\pi^2}{\mu^2} \right) \right]
		-\frac{\D_{ju}^2}{f^2} \ell^{(m)}_{PQ}(\mu)
		+\frac{a^2}{f^2}\ell^{(m)}_{a^2}(\mu)
	\bigg\}\, .
\end{align}

\begin{align}
f_\pi &= f \left\{ 1
	-\frac{2\tilde{m}_{ju}^2}{(4\pi f)^2}\, \ln \left( \frac{\tilde{m}_{ju}^2}{\mu^2} \right)
	+\frac{m_\pi^2}{f^2} \ell^{(f)}(\mu)
	+\frac{\D_{ju}^2}{f^2}\, \ell^{(f)}_{PQ}(\mu)
	+\frac{a^2}{f^2}\, \ell^{(f)}_{a^2}(\mu) \right\} \, .
\end{align}

%%%%%%%%%%%%%%%%%%%%%%%%%%%%%%%%%%%%%%%%%%%%%%%%%%%
%
%		Appendix: meson masses
%
%%%%%%%%%%%%%%%%%%%%%%%%%%%%%%%%%%%%%%%%%%%%%%%%%%%
\section{Meson Masses}
In this Appendix we collect the pion and kaon mass and decay constant for a three-sea flavor MA theory.  These were first computed in Refs.~\cite{Bar:2003mh,Bar:2005tu}.  Here we provide the answers expressed in terms of the PQ parameters we introduced in Eq.~\eqref{eq:PQparameters}.

\begin{align}
m_\pi^2 =&\ 2B_0 \hat{m} \bigg\{ 1 
	+\ln \left( \frac{m_\pi^2}{\mu^2} \right)\, \bigg[
		\frac{m_\pi^2}{(4\pi f)^2}
		-\frac{ \tilde{\D}_{ju}^2 (3 \tilde{m}_{X}^2 -m_\pi^2)}{3(4\pi f)^2 (\tilde{m}_{X}^2 -m_\pi^2)}
		+\frac{\tilde{\D}_{ju}^4 \tilde{m}_X^2}{3(4\pi f)^2 (\tilde{m}_X^2 -m_\pi^2)^2}
		\bigg]
	\nonumber\\ &\qquad
	-\ln \left( \frac{\tilde{m}_X^2}{\mu^2} \right)\, \bigg[
		\frac{\tilde{m}_X^2}{3(4\pi f)^2}
		-\frac{2\tilde{\D}_{ju}^2 \tilde{m}_X^2}{3(4\pi f)^2 (\tilde{m}_X^2 -m_\pi^2)}
		+\frac{\tilde{\D}_{ju}^4 \tilde{m}_X^2}{3(4\pi f)^2 (\tilde{m}_X^2 -m_\pi^2)^2}
		\bigg]
	\nonumber\\ &\qquad
	-\frac{16 m_\pi^2}{f^2}\Big[ L_4(\mu) +L_5(\mu) -2L_6(\mu) -2L_8(\mu) \Big] 
	-\frac{32m_K^2}{f^2} \Big[ L_4(\mu) -2L_6(\mu) \Big] 
	+\frac{a^2}{f^2} L_{ma^2}(\mu)	
	\nonumber\\ &\qquad
	-\left( \frac{32\D_{ju}^2}{f^2} +\frac{16\D_{rs}^2}{f^2} \right) \Big[ L_4(\mu) -2L_6(\mu) \Big] 
	-\frac{\tilde{\D}_{ju}^2}{(4\pi f)^2}
	+\frac{\tilde{\D}_{ju}^4}{3(4\pi f)^2(\tilde{m}_X^2 -m_\pi^2)}
	\bigg\}\, .
\end{align}

\begin{align}
m_K^2 =&\ B_0(\hat{m}+m_s) 
	-\frac{16 m_K^4}{f^2} \Big[ 2 L_4(\mu) +L_5(\mu) -4L_6(\mu) -2L_8(\mu) \Big]
	-\frac{16 m_K^2 m_\pi^2}{f^2} \Big[ L_4(\mu) -2L_6(\mu) \Big]
	\nonumber\\ &\ 
	-\frac{16 m_K^2}{f^2} \Big( 2\tilde{\D}_{ju}^2 +\tilde{\D}_{rs}^2 \Big)
		\Big[ L_4(\mu) -2L_6(\mu) \Big]
	+\frac{a^2 m_K^2}{f^2} L_{ma^2}(\mu)
	\nonumber\\ &\ 
	+\ln \left( \frac{\tilde{m}_X^2}{\mu^2} \right)\, \bigg[
		\frac{2 m_K^2 \tilde{m}_X^2}{3(4\pi f)^2}
		-\frac{\tilde{\D}_{ju}^2 \tilde{m}_X^2 (8m_K^2 +3\tilde{m}_X^2 +m_\pi^2)}
			{18(4\pi f)^2 (\tilde{m}_X^2 -m_\pi^2)}
		+\frac{\tilde{\D}_{ju}^4 \tilde{m}_X^2}{18(4\pi f)^2 (\tilde{m}_X^2 -m_\pi^2)}
	\nonumber\\ &\qquad\qquad\qquad
		-\frac{2\tilde{\D}_{rs}^2 \tilde{m}_X^2 m_K^2}{3(4\pi f)^2 (\tilde{m}_X^2 +m_\pi^2 -2m_K^2)}
		+\frac{\tilde{\D}_{ju}^2 \tilde{\D}_{rs}^2 \tilde{m}_X^2 (\tilde{m}_X^2 +4m_K^2 +m_\pi^2)}
			{9(4\pi f)^2 (\tilde{m}_X^2 -m_\pi^2) (\tilde{m}_X^2 +m_\pi^2 -2m_K^2)}
		\bigg]
	\nonumber\\ &\ 
	+\ln \left( \frac{m_\pi^2}{\mu^2} \right)\, \frac{\tilde{\D}_{ju}^2\, m_\pi^2}{(4\pi f)^2}\, \bigg[
		\frac{3\tilde{m}_X^2 +8m_K^2 +m_\pi^2}{18(\tilde{m}_X^2 -m_\pi^2)}
		-\frac{\tilde{\D}_{ju}^2}{18 (\tilde{m}_X^2 -m_\pi^2)}
		+\frac{\tilde{\D}_{rs}^2 (2m_K^2 +m_\pi^2)}
			{9(m_K^2 -m_\pi^2)(\tilde{m}_X^2 -m_\pi^2)} \bigg]
	\nonumber\\ &\ 
	+\ln \left( \frac{m_{ss}^2}{\mu^2} \right)\, \frac{\tilde{\D}_{rs}^2\, m_K^2}{(4\pi f)^2}\, \bigg[
		\frac{2 (2m_K^2 -m_\pi^2)}{3(\tilde{m}_X^2 +m_\pi^2 -2m_K^2)}
		-\frac{\tilde{\D}_{ju}^2 (2m_K^2 -m_\pi^2)}
			{3(m_K^2 -m_\pi^2)(\tilde{m}_X^2 +m_\pi^2 -2m_K^2)}
	\bigg] .
\end{align}
Note that the lattice spacing dependent counterterms for the meson masses have the same coefficient.  This is because the discretization scheme is flavor-blind.

%%%%%%%%%%%%%%%%%%%%%%%%%%%%%%%%%%%%%%%%%%%%%%%%%%%
%
%		Appendix: meson decay constants
%
%%%%%%%%%%%%%%%%%%%%%%%%%%%%%%%%%%%%%%%%%%%%%%%%%%%
\section{Decay Constants and $f_K / f_\pi$ \label{app:decay_constants}}

The pion decay constant is given by
\begin{align}\label{eq:fpi}
f_\pi &= f \bigg\{ 1
	-\frac{2\tilde{m}_{ju}^2}{(4\pi f)^2}\, \ln \bigg( \frac{\tilde{m}_{ju}^2}{\mu^2} \bigg)
	-\frac{\tilde{m}_{ru}^2}{(4\pi f)^2}\, \ln \bigg( \frac{\tilde{m}_{ru}^2}{\mu^2} \bigg)
	+\frac{8 m_\pi^2}{f^2} \Big( L_5(\mu) +L_4(\mu) \Big)
	+\frac{16 m_K^2}{f^2}\, L_4(\mu) 
	\nonumber\\ &\qquad\qquad 
	+\frac{8(2\D_{ju}^2 +\D_{rs}^2)}{f^2}\, L_4(\mu) 
	+\frac{a^2}{f^2}\, L_{fa^2}(\mu) \bigg\}\, ,
\end{align}
while the kaon decay constant is
\begingroup
%\small
\begin{align}\label{eq:fK}
f_K &= f\bigg\{ 1
	-\frac{\tilde{m}_{sj}^2}{(4 \pi f)^2}\, \ln \bigg( \frac{\tilde{m}_{sj}^2}{\mu^2} \bigg)
	-\frac{\tilde{m}_{ru}^2}{2(4 \pi f)^2}\, \ln \bigg( \frac{\tilde{m}_{ru}^2}{\mu^2} \bigg)
	-\frac{\tilde{m}_{ju}^2}{(4 \pi f)^2}\, \ln \bigg( \frac{\tilde{m}_{ju}^2}{\mu^2} \bigg)
	-\frac{\tilde{m}_{rs}^2}{2(4 \pi f)^2}\, \ln \bigg( \frac{\tilde{m}_{rs}^2}{\mu^2} \bigg)
	+\frac{8 m_\pi^2}{f^2}\, L_4(\mu)
	 \nonumber\\ &
	+\frac{8 m_K^2}{f^2} \Big[ L_5(\mu) +2 L_4(\mu) \Big]
	+\frac{8(2\D_{ju}^2 +\D_{rs}^2)}{f^2}\, L_4(\mu)  
	+\frac{a^2}{f^2}\, L_{fa^2}(\mu) 
	-\frac{\tilde{\D}_{ju}^2}{4(4\pi f)^2}
	+\frac{\tilde{\D}_{ju}^4}{12 (4\pi f)^2 (\tilde{m}_X^2 -m_\pi^2)}
	\nonumber\\ &
	+\frac{\tilde{\D}_{rs}^2 (m_K^2 -m_\pi^2)}{3 (4\pi f)^2 (\tilde{m}_X^2 -m_{ss}^2)}
	-\frac{\tilde{\D}_{ju}^2 \tilde{\D}_{rs}^2}{6(4\pi f)^2 (\tilde{m}_X^2 -m_{ss}^2)}
	+\frac{1}{12(4\pi f)^2} \ln \bigg( \frac{m_\pi^2}{\mu^2} \bigg) \bigg[
		3m_\pi^2
		-\frac{3\tilde{\D}_{ju}^2 (\tilde{m}_X^2 +m_\pi^2)}{\tilde{m}_X^2 -m_\pi^2}
		+\frac{\tilde{\D}_{ju}^4 \tilde{m}_X^2}{(\tilde{m}_X^2 -m_\pi^2)^2}
	\nonumber\\ &
		-\frac{4\tilde{\D}_{ju}^2 \tilde{\D}_{rs}^2 m_\pi^2}
			{(\tilde{m}_X^2 -m_\pi^2)(m_{ss}^2 -m_\pi^2)}
		\bigg]
	-\frac{\tilde{m}_X^2}{12(4\pi f)^2} \ln \bigg( \frac{\tilde{m}_X^2}{\mu^2} \bigg) \bigg[
		9 -\frac{6\tilde{\D}_{ju}^2}{\tilde{m}_X^2 -m_\pi^2}
		+\frac{\tilde{\D}_{ju}^4}{(\tilde{m}_X^2 -m_\pi^2)^2}
	\nonumber\\ &
		+\frac{\tilde{\D}_{rs}^2 \Big( 4(m_K^2 -m_\pi^2) +6(m_{ss}^2 -\tilde{m}_X^2) \Big)}
			{(\tilde{m}_X^2 -m_{ss}^2)^2}
		-\frac{2\tilde{\D}_{ju}^2 \tilde{\D}_{rs}^2 (2m_{ss}^2 -m_\pi^2 -\tilde{m}_X^2)}
			{(\tilde{m}_X^2 -m_{ss}^2)^2 (\tilde{m}_X^2 -m_\pi^2)}
		\bigg]
	+\frac{1}{6(4\pi f)^2} \ln \bigg( \frac{m_{ss}^2}{\mu^2} \bigg) \bigg[
		3m_{ss}^2
	\nonumber\\ &
		+\frac{\tilde{\D}_{rs}^2 \Big( 3m_{ss}^4 +2(m_K^2 -m_\pi^2) \tilde{m}_X^2 
			-3m_{ss}^2 \tilde{m}_X^2 \Big)}{(\tilde{m}_X^2 -m_{ss}^2)^2}
		-\frac{\tilde{\D}_{ju}^2 \tilde{\D}_{rs}^2 (2m_{ss}^4 -\tilde{m}_X^2 (m_{ss}^2 +m_\pi^2)}
			{(\tilde{m}_X^2 -m_{ss}^2)^2 (m_{ss}^2 -m_\pi^2)}
		\bigg]
	\bigg\}\, .
\end{align}
\endgroup

\end{widetext}

The two important things to note are that the additive lattice spacing modifications to the decay constants are the same and also that at this order, they can be absorbed into a redefinition of the Lagrangian parameter, $f$.  We can then use these formulae to estimate the size of the corrections to the recent determination of $L_5(\mu)$ by NPLQCD~\cite{Beane:2006kx}.  Thus, we form the ratio
\begingroup
\small
\begin{equation}
	\Delta \left( \frac{f_K}{f_\pi} \right) = 
	\frac{ \frac{f_K}{f_\pi} \bigg|_{MA} - \frac{f_K}{f_\pi} \bigg|_{QCD}}{\frac{f_K}{f_\pi} \bigg|_{QCD}}\, ,
\end{equation}
\endgroup
where, using Eqs.~\eqref{eq:fpi} and \eqref{eq:fK}, and the tuning used in Ref.~\cite{Beane:2006kx} which was to set the valence-valence meson masses equal to the taste-$\xi_5$ sea-sea mesons, we have
\begin{widetext}
\begingroup
\small
\begin{align}\label{eq:fKfpiMAQCD}
\frac{f_K}{f_\pi} \bigg|_{MA} - \frac{f_K}{f_\pi} \bigg|_{QCD} &=  
	\frac{m_\pi^2 +a^2 \D_\mathrm{Mix}}{(4\pi f_\pi)^2}\, 
			\ln \bigg( \frac{m_\pi^2 +a^2 \D_\mathrm{Mix}}{\mu^2} \bigg) 
	-\frac{m_\pi^2}{(4\pi f_\pi)^2}\, \ln \bigg( \frac{m_\pi^2}{\mu^2} \bigg) 
	-\frac{m_K^2 +a^2 \D_\mathrm{Mix}}{2(4\pi f_\pi)^2}\, 
		\ln \bigg( \frac{m_K^2 +a^2 \D_\mathrm{Mix}}{\mu^2} \bigg) 
\nonumber\\ &
	+\frac{m_K^2}{2(4\pi f_\pi)^2}\, \ln \bigg( \frac{m_K^2}{\mu^2} \bigg)
	-\frac{3}{4(4\pi)^2} \bigg[
		\frac{m_\eta^2 +a^2 \D_{I}}{f_\pi^2}\, \ln \bigg( \frac{m_\eta^2 +a^2 \D_{I}}{\mu^2} \bigg)
		-\frac{m_\eta^2}{f_\pi^2}\, \ln \bigg( \frac{m_\eta^2}{\mu^2} \bigg) \bigg] 
	\nonumber\\ & 
	-\frac{1}{2(4\pi)^2} \left[ 
		\frac{m_{ss}^2 +a^2 \D_\mathrm{Mix}}{f_\pi^2}\, 
			\ln \bigg( \frac{m_{ss}^2 +a^2 \D_\mathrm{Mix}}{\mu^2} \bigg) 
		- \frac{m_{ss}^2}{f_\pi^2}\, \ln \bigg( \frac{m_{ss}^2}{\mu^2} \bigg) \right] 
	\nonumber\\ & 
	-\bigg( \frac{a^2 \D_{I}}{f_\pi^2} \bigg) \frac{1}{12(4\pi)^2} \bigg\{
	3 +\frac{4(m_K^2 -m_\pi^2)}{m_{ss}^2 -m_{\eta}^2 -a^2 \D_I}
	+\frac{3(m_{\eta}^2 +a^2 \D_I +m_\pi^2)}{m_{\eta}^2 +a^2 \D_I -m_\pi^2}\, 
		\ln \bigg( \frac{m_{\pi}^2}{\mu^2} \bigg) 
	\nonumber\\ &  
	-\frac{2\Big(3 m_{ss}^4 -(m_{\eta}^2 +a^2\D_I) (3m_{ss}^2 -2m_K^2 +2m_\pi^2 ) \Big)}
		{(m_{ss}^2 -m_{\eta}^2 -a^2\D_I)^2}\, \ln \bigg( \frac{m_{ss}^2}{\mu^2} \bigg)
	\nonumber\\ &  
	+2(m_{\eta}^2 +a^2\D_I)\, \ln \bigg( \frac{m_{\eta}^2 +a^2\D_I}{\mu^2} \bigg)
	\bigg[ \frac{3m_{ss}^2 -3(m_{\eta}^2 +a^2\D_I) +2m_K^2 -2m_\pi^2}
		{(m_{ss}^2 -m_{\eta}^2 -a^2\D_I)^2}
	-\frac{3}{m_{\eta}^2 +a^2\D_I -m_\pi^2}
	\bigg]
	\bigg\}
	\nonumber\\ & 
	+\bigg( \frac{a^2 \D_{I}}{f_\pi^2} \bigg)^2 \frac{1}{12(4\pi)^2} \bigg\{
		\frac{f_\pi^2}{m_\eta^2 +a^2 \D_{I}-m_\pi^2} 
		-\frac{2 f_\pi^2}{m_{\eta}^2 +a^2 \D_I -m_{ss}^2}
	+\ln \bigg( \frac{m_\pi^2}{\mu^2} \bigg)
	\bigg[ \frac{f_\pi^2 (m_{\eta}^2 +a^2 \D_I)}{(m_\eta^2 +a^2 \D_{I}-m_\pi^2)^2}
	\nonumber\\ & 
		-\frac{4 f_\pi^2 m_\pi^2}{(m_\eta^2 +a^2 \D_{I}-m_\pi^2)(m_{ss}^2 -m_\pi^2)}
	\bigg]
	-\ln \bigg( \frac{m_{ss}^2}{\mu^2} \bigg)
	\frac{2f_\pi^2 \Big( 2m_{ss}^4 -(m_{\eta}^2 +a^2\D_I)(m_{ss}^2 +m_\pi^2) \Big)}
		{(m_{ss}^2 -m_\pi^2)(m_{\eta}^2 +a^2\D_I -m_{ss}^2)^2}\,
	\nonumber\\ & 
	-\frac{m_{\eta}^2 +a^2 \D_I}{f_\pi^2}\, \ln \bigg( \frac{m_{\eta}^2 +a^2 \D_I}{\mu^2} \bigg)
	\bigg[ \frac{f_\pi^4}{(m_{\eta}^2 +a^2 \D_I -m_\pi^2)^2} 
		+\frac{2f_\pi^4 (m_{\eta}^2 +a^2 \D_I +m_\pi^2 -2m_{ss}^2)}
			{(m_{\eta}^2 +a^2 \D_I -m_\pi^2)(m_{\eta}^2 +a^2 \D_I -m_{ss}^2)^2} \bigg]
	\bigg\}
\end{align}
\endgroup

\end{widetext}

%%%%%%%%%%%%%%%%%%%%%%%%%%%%%%%%%%%%%%%%%%%%%%%%%%%
%
%		Appendix: pi-pi scattering
%
%%%%%%%%%%%%%%%%%%%%%%%%%%%%%%%%%%%%%%%%%%%%%%%%%%%
\section{$\pi^+\pi^+$ Scattering \label{app:pipi}}

For completeness, we provide the formulae for the $I=2\ \pi\pi$ scattering length determined in both MA$\chi$PT and PQ$\chi$PT in Ref.~\cite{Chen:2005ab}, for both two and three flavors of sea quark.

\subsection*{Two sea Quark Flavors, $m_\pi a_{\pi\pi}^{I=2}$}
\begin{multline}\label{eq:I2pipiSU2}
m_\pi a_{\pi\pi}^{I=2} = \frac{-m_\pi^2}{8 \pi f_\pi^2} \bigg\{ 1
	+\frac{m_\pi^2}{(4 \pi f_\pi)^2}\, \bigg[
		3 \ln \left( \frac{m_\pi^2}{\mu^2} \right) -1
	\\
		-l_{\pi\pi}^{I=2}(\mu)
		\bigg]
		-\frac{m_\pi^2}{(4\pi f_\pi)^2}\, \frac{\tilde{\D}_{ju}^4}{6 m_\pi^4}
	\bigg\}
\end{multline}

\subsection*{Three sea Quark Flavors, $m_\pi a_{\pi\pi}^{I=2}$}
\begin{multline}\label{eq:I2pipiSU3}
m_\pi a_{\pi\pi}^{I=2} = \frac{-m_\pi^2}{8 \pi f_\pi^2} \bigg\{ 1
	+\frac{m_\pi^2}{(4 \pi f_\pi)^2}\, \bigg[
		3 \ln \left( \frac{m_\pi^2}{\mu^2} \right) -1
	\\
	+\frac{1}{9}\ln \left( \frac{\tilde{m}_X^2}{\mu^2} \right) +\frac{1}{9}
	-32 (4\pi)^2\, L_{\pi\pi}^{I=2}(\mu)
	\bigg] \\
	+\frac{m_\pi^2}{(4\pi f_\pi)^2} \bigg[ -\frac{\tilde{\D}_{ju}^4}{6 m_\pi^4}
	\\
	+\sum_{n=1}^4 \left( \frac{\tilde{\D}_{ju}^2}{m_\pi^2} \right)^n\, 
		\mc{F}_n (m_\pi^2 / \tilde{m}_X^2 )
	\bigg] \bigg\}\, ,
\end{multline}
where the functions $\mc{F}_n(y)$ are given by

\begin{widetext}
\begin{subequations}
\begin{align}
        \mc{F}_1(y) &= -\frac{2y}{9(1-y)^2} \Big[ 5(1-y) +(3 +2y)\ln (y) \Big], \\ %\nonumber\\
        \mc{F}_2(y) &= \frac{2y}{3(1-y)^3} \Big[ (1-y)(1+3y) +y(3 +y)\ln(y) \Big], \\ %\nonumber\\
        \mc{F}_3(y) &= \frac{y}{9(1-y)^4} \Big[ (1-y) (1 -7y -12y^2) -2y^2(7+2y) \ln(y) \Big], \\ %\nonumber\\
        \mc{F}_4(y) &= -\frac{y^2}{54 (1-y)^5} \Big[ (1-y) (1 -8y -17y^2) -6y^2(3+y)\ln(y) \Big]\, .
\end{align}\label{eq:coolFs}\end{subequations}

\end{widetext}
%

%%%%%%%%%%%%%%%%%%%%%%%%%%%%%%%%%%%%%%%%%%%%%%%%%%%
%
%	Applications: K-K scattering
%
%%%%%%%%%%%%%%%%%%%%%%%%%%%%%%%%%%%%%%%%%%%%%%%%%%%
\section{$K^+K^+$ Scattering}\label{app:kk}
The $I=1\ KK$ and $I=3/2\ K\pi$ scattering lengths involve lengthy expressions.  Therefore, we introduce the following notation to make the answers more presentable.
\begin{equation}\label{eq:kjX}
	m_K = k m_\pi \quad,\quad
	\tilde{\D}_{ju} = \d_{ju} m_\pi \quad,\quad
	\tilde{\D}_{rs} = \d_{rs} m_\pi \, .
\end{equation}

The $I=1\ KK$ scattering length is given by Eq.~\eqref{eq:aKK}, which we repeat here for convenience
\begin{multline*}
m_K a_{KK}^{I=1} =-\frac{m_{K}^{2}}{8 \pi f_{K}^{2}} \bigg\{ 1
	+\frac{m_{K}^{2}}{ (4\pi f_{K})^{2}} \bigg[ 
		C_\pi \ln \left( \frac{m_\pi^{2}}{\mu^2} \right) 
	\\
		+C_K \ln \left( \frac{m_K^{2}\,}{\mu ^2} \right) 
		+C_X \ln \left( \frac{\tilde{m}_X^{2}}{\mu^2} \right) 
	\\
		+C_{ss} \ln \left( \frac{m_{ss}^{2}}{\mu ^{2}} \right) 
		+ C_0 
	- 32(4\pi)^2\, L_{KK}^{I=1}
	\bigg] \bigg\}\, ,
\end{multline*}
where

\begin{equation}
C_K = 2\, ,
\end{equation}

\begin{align}\label{eq:KKCpi}
C_\pi =& \frac{2 -2 k^2 -\d_{rs}^2}{(k^2 -1)^3 (4 k^2 - 4 + \delta_{ju}^2 + 2 \delta_{rs}^2)^3}
	\big\{ 
	16 (k^2-1)^4 
	\nonumber\\&
	+16 (k^2-1)^3 \d_{rs}^2 + 4(k^2-1)^2 \d_{rs}^4 
	+\d_{ju}^2 \big[ 4(k^2-1)^3
	\nonumber\\&
	\times ( 5 + 4k^2) +2(k^2-1)^2 (5 +8k^2)\d_{rs}^2 
	\nonumber\\&
	+4k^2 ( k^2-1)\d_{rs}^4 \big]
	-\d_{ju}^4 \big[ 4 (k^2-1)^2 
	\nonumber\\&
	+4( k^4-1)\d_{rs}^2 +2 k^2 \d_{rs}^4 \big] 
	\nonumber\\&
	-\d_{ju}^6 \big[ (k^2-1)^2 +k^2 \d_{rs}^2 \big] 
	\big\}\, ,
\end{align}

% widetext version of equation
\begin{comment}
\begin{align}\label{eq:KKCpi}
C_\pi =&\ \frac{2 -2 k^2 -\d_{rs}^2}{(k^2 -1)^3 (4 k^2 - 4 + \delta_{ju}^2 + 2 \delta_{rs}^2)^3}
	\bigg[ 
	16 (k^2-1)^4 + 16 (k^2-1)^3 \d_{rs}^2 + 4(k^2-1)^2 \d_{rs}^4 
	\nonumber\\ &\qquad
	+\d_{ju}^2 \Big( 4(k^2-1)^3( 5 + 4k^2) +2(k^2-1)^2 (5 +8k^2)\d_{rs}^2 +4k^2 ( k^2-1)\d_{rs}^4 \Big)
	\nonumber\\ &\qquad
	-\d_{ju}^4 \Big( 4 (k^2-1)^2 +4( k^4-1)\d_{rs}^2 +2 k^2 \d_{rs}^4 \Big) 
	-\d_{ju}^6 \Big( (k^2-1)^2 +k^2 \d_{rs}^2 \Big) 
	\bigg]\, ,
\end{align}
\end{comment}
%
\begin{widetext}

\begin{align}
C_X =& -\frac{8 (2 -2 k^2 + \d_{ju}^2 - \d_{rs}^2)^2}
	{9(2 -2 k^2 +\d_{ju}^2 + 2\d_{rs}^2)^3 ( 4k^2 -4 +\d_{ju}^2 +2\d_{rs}^2)^3}
	\Big\{
	8 (k^2-1)^3 (20 k^2 -11)  
	\nonumber\\ &
	+4(k^2-1)^2 ( 152 k^2 -53 ) \d_{rs}^2 
	+12 ( 38 k^4 - 61 k^2 +23) \d_{rs}^4
	+80 (k^2 -1)\d_{rs}^6 
	-8\d_{rs}^8 
	\nonumber\\ &
	+\d_{ju}^2 \Big[ 14(k^2-1)^2 (1 +8k^2) -24( 5k^4 -4 -k^2 )\d_{rs}^2 
		-( 312k^2 -132)\d_{rs}^4 -112 \d_{rs}^6 \Big]
	\nonumber\\ &
	-\d_{ju}^4 \Big[ 33 ( 2k^4 -k^2 -1) +( 210 k^2 -138 )\d_{rs}^2 +102\d_{rs}^4 \Big] 
	-\d_{ju}^6 \Big[ 17 k^2 -26  +22 \d_{rs}^2 \Big] 
	+\d_{ju}^8 
	\Big\}\, ,
\end{align}
% different format of above equation
\begin{comment}
\begin{align}
C_X =&\ -\frac{8 (2 -2 k^2 + \d_{ju}^2 - \d_{rs}^2)^2}
	{9(2 -2 k^2 +\d_{ju}^2 + 2\d_{rs}^2)^3 ( 4k^2 -4 +\d_{ju}^2 +2\d_{rs}^2)^3}
	\bigg[
	8 (k^2-1)^3 (20 k^2 -11)  
	\nonumber\\ &\qquad
	+4(k^2-1)^2 ( 152 k^2 -53 ) \d_{rs}^2 
	+12 ( 38 k^4 - 61 k^2 +23) \d_{rs}^4
	+80 (k^2 -1)\d_{rs}^6 
	-8\d_{rs}^8 
	\nonumber\\ &\qquad
	+\d_{ju}^2 \Big( 14(k^2-1)^2 (1 +8k^2) -24( 5k^4 -4 -k^2 )\d_{rs}^2 
		-( 312k^2 -132)\d_{rs}^4 -112 \d_{rs}^6 \Big)
	\nonumber\\ &\qquad
	-\d_{ju}^4 \Big( 33 ( 2k^4 -k^2 -1) +( 210 k^2 -138 )\d_{rs}^2 +102\d_{rs}^4 \Big) 
	-\d_{ju}^6 \Big( 17 k^2 -26  +22 \d_{rs}^2 \Big) 
	+\d_{ju}^8 
	\bigg]\, ,
\end{align}
\end{comment}

\begin{align}
C_{ss} =&\ \frac{\d_{rs}^2}{(k^2-1)^3 ( 2 -2k^2 + \d_{ju}^2 +2 \d_{rs}^2)^3}
	\Big\{ 
	8 (k^2-1)^4 (7 k^2 -3 ) 
	-4 (k^2-1)^3 (4 k^2 -1 )  \d_{rs}^2 
	+4 (k^2-1)^2 (2 k^2 -1 )  \d_{rs}^4 
	\nonumber\\ &
	-\d_{ju}^2 \Big[ 4 (k^2-1)^3 ( 17 k^2 -7)  + 2 (k^2-1)^2(4k^2 -3)\d_{rs}^2 
		-4 k^2 (k^2 -1 )  \d_{rs}^4 \Big]
	-\d_{ju}^6 \Big[ 3 k^4 -4k^2 +1 + k^2 \d_{rs}^2 \Big] 
	\nonumber\\ &
	+\d_{ju}^4 \Big[ 2 (k^2-1)^2 (13 k^2 -5 ) +2(5k^4 -7k^2 +2 )\d_{rs}^2 -2k^2\d_{rs}^4 \Big] 
	\Big\} \, ,
\end{align}

% different format of above equation
\begin{comment}
\begin{align}
C_{ss} =&\ \frac{\d_{rs}^2}{(k^2-1)^3 ( 2 -2k^2 + \d_{ju}^2 +2 \d_{rs}^2)^3}
	\bigg[ 
	8 (k^2-1)^4 (7 k^2 -3 ) 
	-4 (k^2-1)^3 (4 k^2 -1 )  \d_{rs}^2 
	\nonumber\\ &\quad
	+4 (k^2-1)^2 (2 k^2 -1 )  \d_{rs}^4 
%	\nonumber\\ &\quad
	-\d_{ju}^2 \Big( 4 (k^2-1)^3 ( 17 k^2 -7)  + 2 (k^2-1)^2(4k^2 -3)\d_{rs}^2 
		-4 k^2 (k^2 -1 )  \d_{rs}^4 \Big)
	\nonumber\\ &\quad
	+\d_{ju}^4 \Big( 2 (k^2-1)^2 (13 k^2 -5 ) +2(5k^4 -7k^2 +2 )\d_{rs}^2 -2k^2\d_{rs}^4 \Big) 
	-\d_{ju}^6 \Big( 3 k^4 -4k^2 +1 + k^2 \d_{rs}^2 \Big) 
	\bigg] \, ,
\end{align}
\end{comment}

\begin{align}\label{eq:KKC0}
C_0 =&\ \frac{2}{9 (k^2 -1)^2 (4 k^2 -4 +\d_{ju}^2 + 2\d_{rs}^2)^2 (2 -2k^2 +\d_{ju}^2 +2\d_{rs}^2)^2}
	\Big\{
	-448 (k^2-1)^6 
	+1120 (k^2-1)^5 \d_{rs}^2 
	\nonumber\\ &
	+912 (k^2-1)^4 \d_{rs}^4 
	-152 (k^2-1)^3 \d_{rs}^6 
	-136 (k^2-1)^2 \d_{rs}^8 
	+\d_{ju}^8 \Big[ 8 (k^2-1)^2 +18 (k^2-1)\d_{rs}^2 + 9 \d_{rs}^4 \Big] 
	\nonumber\\ &
	-\d_{ju}^2 \Big[ 112 (k^2-1)^5 
	-48 (k^2-1)^4 \d_{rs}^2 
	+876 (k^2-1)^3 \d_{rs}^4 
	+608 (k^2-1)^2 \d_{rs}^6 +72 (k^2-1)  \d_{rs}^8 \Big]
	\nonumber\\ &
	+\d_{ju}^4 \Big[ 480 (k^2-1)^4 -96 (k^2-1)^3 \d_{rs}^2 
	-330 (k^2-1)^2 \d_{rs}^4 
	+36 (k^2-1)  \d_{rs}^6 + 36 \d_{rs}^8 \Big] 
	\nonumber\\ &
	-\d_{ju}^6 \Big[ 172 (k^2-1)^3 +140 (k^2-1)^2 \d_{rs}^2 -72 (k^2-1)\d_{rs}^4 -36 \d_{rs}^6 \Big] 
	\Big\} \, .
\end{align}

% different format of above equation
\begin{comment}
\begin{align}\label{eq:KKC0}
C_0 =&\ \frac{2}{9 (k^2 -1)^2 (4 k^2 -4 +\d_{ju}^2 + 2\d_{rs}^2)^2 (2 -2k^2 +\d_{ju}^2 +2\d_{rs}^2)^2}
	\bigg[
	-448 (k^2-1)^6 
	+1120 (k^2-1)^5 \d_{rs}^2 
	\nonumber\\ &\quad
	+912 (k^2-1)^4 \d_{rs}^4 
	-152 (k^2-1)^3 \d_{rs}^6 
	-136 (k^2-1)^2 \d_{rs}^8 
	+\d_{ju}^8 \Big( 8 (k^2-1)^2 +18 (k^2-1)\d_{rs}^2 + 9 \d_{rs}^4 \Big) 
	\nonumber\\ &\quad
	-\d_{ju}^2 \Big( 112 (k^2-1)^5 -48 (k^2-1)^4 \d_{rs}^2 +876 (k^2-1)^3 \d_{rs}^4 
		+608 (k^2-1)^2 \d_{rs}^6 +72 (k^2-1)  \d_{rs}^8 \Big)
	\nonumber\\ &\quad
	+\d_{ju}^4 \Big( 480 (k^2-1)^4 -96 (k^2-1)^3 \d_{rs}^2 -330 (k^2-1)^2 \d_{rs}^4 
		+36 (k^2-1)  \d_{rs}^6 + 36 \d_{rs}^8 \Big) 
	\nonumber\\ &\quad
	-\d_{ju}^6 \Big( 172 (k^2-1)^3 +140 (k^2-1)^2 \d_{rs}^2 -72 (k^2-1)\d_{rs}^4 -36 \d_{rs}^6 \Big) 
	\bigg] \, .
\end{align}
\end{comment}

%\end{widetext}

%%%%%%%%%%%%%%%%%%%%%%%%%%%%%%%%%%%%%%%%%%%%%%%%%%%
%
%	Applications: K-pi scattering
%
%%%%%%%%%%%%%%%%%%%%%%%%%%%%%%%%%%%%%%%%%%%%%%%%%%%
\section{$K^+\pi^+$ Scattering}\label{app:kpi}
%
%\begin{widetext}
%

The $K\pi$ scattering length at $I=3/2$ is given by:
\begin{equation*}
\mu_{K\pi} a_{K\pi}^{I=3/2} = -\frac{\mu_{K\pi}^2}{4\pi f_K f_\pi}\, \bigg[
	1 -\frac{32 m_Km_\pi}{f_K f_\pi}\, L_{\pi\pi}^{I=2}(\mu)
	+\frac{8(m_K -m_\pi)^2}{f_K f_\pi}\, L_5(\mu) \bigg]
	+\mu_{K\pi} \Big[ a_{vv}^{K\pi,3/2}(\mu) +a_{vs}^{K\pi,3/2}(\mu) \Big]\, ,
\end{equation*}
where $a_{vs}^{K\pi,3/2}(\mu)$ is given in Eq.~\eqref{eq:aKpiVS}. %
%\begin{comment}
\begin{equation*}
\mu_{K\pi}\, a_{vs}^{K\pi,3/2}(\mu) = -\frac{\mu_{K\pi}^2}{4\pi f_K f_\pi} \frac{1}{2(4\pi)^2 f_K f_\pi}
	\sum_{F=j,l,r} 
	\bigg[ C_{Fs} \ln \left( \frac{\tilde{m}_{Fs}^2}{\mu^2} \right)
	-C_{Fd} \ln \left( \frac{\tilde{m}_{Fd}^2}{\mu^2} \right)
	+4m_K m_\pi J(\tilde{m}_{Fd}^2)
	\bigg]\, ,
\end{equation*}
and the coefficients, $C_{Fd,s}$ and the function, $J(m)$ are defined in Eqs.~\eqref{eq:CFs}--\eqref{eq:JVS}.  
We reiterate that the $\ln(\mu^2)$ dependence in $a_{vs}^{K\pi,3/2}(\mu)$ only depends upon the valence-valence meson masses, Eq.~\eqref{eq:a32VSmu}, as we argued in Sec.~\ref{sec:NLO}.  
%\end{comment}
The valence-valence (and valence-ghost) contribution to the scattering length is given by
\begin{multline}
\mu_{K\pi} a_{vv}^{K\pi,3/2}(\mu) = \frac{\mu_{K\pi}^2}{4\pi f_K f_\pi} \frac{m_\pi^2}{2(4\pi)^2 f_K f_\pi} \bigg[
	A_\pi\, \ln \left( \frac{m_\pi^2}{\mu^2} \right) 
	+A_K\, \ln \left( \frac{m_K^2}{\mu^2} \right) 
	\\
	+A_X\, \ln \left( \frac{\tilde{m}_X^2}{\mu^2} \right) 
	+A_{ss}\, \ln \left( \frac{m_{ss}^2}{\mu^2} \right) 
	+A_{tan} 
	+A_0
	\bigg]\, .
\end{multline}
We use the notation defined in Eq.~\eqref{eq:kjX} to simplify the form of these coefficients. We find

\begin{align}
A_\pi =&\ \frac{1}{(k^2 -1)^3 (4k^2 -4 + \d_{ju}^2 +2 \d_{rs}^2)^4 } \Big\{ 
	8{( k^2 -1 ) }^2 ( 14k -k^2 -1) ( 2k^2 -2 +\d_{rs}^2 )^4 
	+8\d_{ju}^2( k^2 -1 ) ( 2k^2 -2 +\d_{rs}^2)^3 
	\nonumber\\ & \times
	\Big[ 2k^6 +k^4( \d_{rs}^2 -1) + 28k^3 
	+k^2( \d_{rs}^2 -2) -k( 28 -\d_{rs}^2 ) +1 \Big]  
	+2\d_{ju}^4 ( 2k^2 -2 +\d_{rs}^2 )^2 \Big[ 12k^8 +48k^7 
	\nonumber\\ &
	+3k^6( -5 +2\d_{rs}^2 ) +32k^5\d_{rs}^2 -9k^4 
	-2k^3( 72 +21\d_{rs}^2 -2\d_{rs}^4) +k^2( 15 -6\d_{rs}^2) 
	+2k( 48 +5\d_{rs}^2 -\d_{rs}^4) -3 \Big] 
	\nonumber\\ &
	+2\d_{ju}^6 \Big[ 4(k^2 -1)^3 (3k^4 +12k^3+2k^2+5k-1)
	+2 \d_{rs}^2 (k^2-1)^2 (6k^4 +32k^3 +5k^2 -2k -1) 
	\nonumber\\ &
	+k \d_{rs}^4 (3k^5 +28k^4 -39k^2 -3k +11)
	+2k \d_{rs}^6 (2 k^2-1) \Big] 
	+\d_{ju}^8 \Big[ 2k^8 -k^6( 3 -\d_{rs}^2 ) +2k^5( 5 +2\d_{rs}^2 ) 
	\nonumber\\ &
	-k^4 -k^3( 20 +5\d_{rs}^2 -2\d_{rs}^4) 
	+k^2(3- \d_{rs}^2 ) +k( 10 + \d_{rs}^2 - \d_{rs}^4 )  -1 \Big]  
	\Big\} \, ,
\end{align}

% different format of above equation
\begin{comment}
\begin{align}
A_\pi =&\ \frac{1}{(k^2 -1)^3 (4k^2 -4 + \d_{ju}^2 +2 \d_{rs}^2)^4 } \bigg[ 
	8{( k^2 -1 ) }^2 ( 14k -k^2 -1) ( 2k^2 -2 +\d_{rs}^2 )^4 
	\nonumber\\ &
	+8\d_{ju}^2( k^2 -1 ) ( 2k^2 -2 +\d_{rs}^2)^3 \Big( 2k^6 +k^4( \d_{rs}^2 -1) + 28k^3 
		+k^2( \d_{rs}^2 -2) -k( 28 -\d_{rs}^2 ) +1 \Big)  
	\nonumber\\ &
	+2\d_{ju}^4 ( 2k^2 -2 +\d_{rs}^2 )^2 \Big( 12k^8 +48k^7 +3k^6( -5 +2\d_{rs}^2 ) +32k^5\d_{rs}^2 
		-9k^4 
	\nonumber\\ &\qquad
		-2k^3( 72 +21\d_{rs}^2 -2\d_{rs}^4) +k^2( 15 -6\d_{rs}^2) 
		+2k( 48 +5\d_{rs}^2 -\d_{rs}^4) -3 \Big) 
	\nonumber\\ &
	+2\d_{ju}^6 \Big( 4(k^2 -1)^3 (3k^4 +12k^3+2k^2+5k-1)
		+2 \d_{rs}^2 (k^2-1)^2 (6k^4 +32k^3 +5k^2 -2k -1) 
	\nonumber\\ &\qquad
		+k \d_{rs}^4 (3k^5 +28k^4 -39k^2 -3k +11)
		+2k \d_{rs}^6 (2 k^2-1)
		\Big) 
	\nonumber\\ &
	+\d_{ju}^8 \Big( 2k^8 -k^6( 3 -\d_{rs}^2 ) +2k^5( 5 +2\d_{rs}^2 ) -k^4 
		-k^3( 20 +5\d_{rs}^2 -2\d_{rs}^4) 
	\nonumber\\ &\qquad
		+k^2(3- \d_{rs}^2 ) +k( 10 + \d_{rs}^2 - \d_{rs}^4 )  -1 \Big)  
	\bigg] \, ,
\end{align}
\end{comment}

\begin{equation}
A_K = \frac{-2k}{9 (k-1)^2 (k+1)} \bigg[
	40k^3 - 26k^2 - 4k -10 -( 1 + k )(2\delta_{ju}^2 +\d_{rs}^2) \bigg] \, ,
\end{equation}

\begin{align}
A_{ss}=&\ -\frac{1}{(k^2 -1)^3 (2-2k^2 + \d_{ju}^2 + 2\d_{rs}^2)^2} \Big\{
	2(k -1)^2 (k+1)^3 \Big[ 4k^7 -4k^6 +2k^5( -5 +\d_{rs}^2 ) 
	+2k^4( 3 + 5\d_{rs}^2) 
 	-2 -\d_{rs}^4
 	\nonumber\\ &
	+2k^3( 4 -\d_{rs}^2 ) 
	- 8k^2( \d_{rs}^2 + \d_{rs}^4 )  - k( 2 + 2\d_{rs}^2 + 5\d_{rs}^4 ) \Big] 
	-2\d_{ju}^2(k^2 -1) \Big[ 4k^8 +2k^6( -7 + \d_{rs}^2 ) 
	+4k^5( -1 + 3\d_{rs}^2)
	\nonumber\\ &
	+k^4( 14 + 8\d_{rs}^2 - \d_{rs}^4 )  
	-2k^3( -4 + 5\d_{rs}^2 + \d_{rs}^4) -k^2( 2 + 10\d_{rs}^2 + 5\d_{rs}^4 )
	-k( 4 + 2\d_{rs}^2 + 3\d_{rs}^4 )  -2 \Big]
	\nonumber\\ &
	+\d_{ju}^4 \Big[ 2k^8 -k^6(7 -\d_{rs}^2 ) -k^5( 2 -6\d_{rs}^2) +k^4( 7 + 4\d_{rs}^2)  
	-k^3( -4 + 5\d_{rs}^2 - 2\d_{rs}^4 ) 
	\nonumber\\ &
	-k^2( 1 + 5\d_{rs}^2 ) -k( 2 + \d_{rs}^2 + \d_{rs}^4 ) -1\Big] 
	\Big\} \, ,
\end{align}

% different format of above equation
\begin{comment}
\begin{align}
A_{ss}=&\ -\frac{1}{(k^2 -1)^3 (2-2k^2 + \d_{ju}^2 + 2\d_{rs}^2)^2} \bigg[
	2(k -1)^2 (k+1)^3 \Big( 4k^7 -4k^6 +2k^5( -5 +\d_{rs}^2 ) 
	\nonumber\\ &\qquad
		+2k^4( 3 + 5\d_{rs}^2) +2k^3( 4 -\d_{rs}^2 ) 
		- 8k^2( \d_{rs}^2 + \d_{rs}^4 )  - k( 2 + 2\d_{rs}^2 + 5\d_{rs}^4 ) -2 -\d_{rs}^4 \Big) 
	\nonumber\\ &
	-2\d_{ju}^2(k^2 -1) \Big( 4k^8 +2k^6( -7 + \d_{rs}^2 ) +4k^5( -1 + 3\d_{rs}^2)
		+k^4( 14 + 8\d_{rs}^2 - \d_{rs}^4 )  
	\nonumber\\ &\qquad
	-2k^3( -4 + 5\d_{rs}^2 + \d_{rs}^4) -k^2( 2 + 10\d_{rs}^2 + 5\d_{rs}^4 )
	-k( 4 + 2\d_{rs}^2 + 3\d_{rs}^4 )  -2 \Big)
	\nonumber\\ &
	+\d_{ju}^4 \Big( 2k^8 -k^6(7 -\d_{rs}^2 ) -k^5( 2 -6\d_{rs}^2) +k^4( 7 + 4\d_{rs}^2)  
		-k^3( -4 + 5\d_{rs}^2 - 2\d_{rs}^4 ) 
	\nonumber\\ &\qquad
		-k^2( 1 + 5\d_{rs}^2 ) -k( 2 + \d_{rs}^2 + \d_{rs}^4 ) -1\Big) 
	\bigg] \, ,
\end{align}
\end{comment}

\begin{align}
A_X =&\ \frac{4(2-2k^2 +\d_{ju}^2 -\d_{rs}^2)^2}
	{9(k-1)^2 (2-2k^2 +\d_{ju}^2 +2\d_{rs}^2)^2 (4 k^2 -4 +\d_{ju}^2 +2 \d_{rs}^2)^4} \Big\{
	-\d_{ju}^{10}k   - 2k \d_{ju}^8 \Big[ 5k^2 +7k -12 +5\d_{rs}^2 \Big] 
	\nonumber\\ &
	+2\d_{ju}^6 \Big[ 9k^6 -126k^5 +113k^4 + k^3( 139 -100\d_{rs}^2 )  
	+k^2( -167 + 64\d_{rs}^2 ) +k( 41 + 36\d_{rs}^2 - 20\d_{rs}^4 ) -9 \Big]  
	\nonumber\\ &
	+2\d_{ju}^4 \Big[ 108k^8 -488k^7 +3k^6( 73 + 18\d_{rs}^2 ) +k^5( 570 - 828\d_{rs}^2 ) 
	+3k^4( -47 + 274\d_{rs}^2 ) +81-54\d_{rs}^2 
	\nonumber\\ &
	-6k^3( 36 -97\d_{rs}^2 +62\d_{rs}^4 )  +3k^2( -89 - 214\d_{rs}^2 + 112\d_{rs}^4 ) 
	+k( 134 + 66\d_{rs}^2 + 36\d_{rs}^4 - 40\d_{rs}^6) \Big]  
	\nonumber\\ &
	+8\d_{ju}^2 \Big[ 108k^{10} -56k^9 +2k^8( -251 +54\d_{rs}^2 ) -4k^7( -87 +86\d_{rs}^2 ) 
	+3k^6( 232 -23\d_{rs}^2 +9\d_{rs}^4)   
	\nonumber\\ &
	-6k^5( 82 -104\d_{rs}^2 +57\d_{rs}^4) +3k^4( -124 +13\d_{rs}^2 +89\d_{rs}^4) 
	+k^3( 164 -360\d_{rs}^2 +273\d_{rs}^4 -112\d_{rs}^6) 
	\nonumber\\ &
	+k^2( 124 -159\d_{rs}^2 -141\d_{rs}^4 + 88\d_{rs}^6 )  
	+k( 36 +80\d_{rs}^2 -57\d_{rs}^4 +24\d_{rs}^6 -10\d_{rs}^8 ) 
	-27( 2 -3\d_{rs}^2 +\d_{rs}^4 )\Big]
	\nonumber\\ &
	+8(2k^2 -2 + \d_{rs}^2 )^2 \Big[ 36k^8 -24k^7 +k^6( -85 +18\d_{rs}^2) +k^5( 40 - 28\d_{rs}^2 ) 
		+k^4( 71 -46\d_{rs}^2)  +9
	\nonumber\\ &
		-2k^3( 4 - 33\d_{rs}^2 + 8\d_{rs}^4 ) +k^2( -31 + 10\d_{rs}^2 - 4\d_{rs}^4 )   
		-2k( 4 + \d_{rs}^2 - 10\d_{rs}^4 + 2\d_{rs}^6 ) -18\d_{rs}^2 \Big]
	\Big\} \, ,
\end{align}

\begin{align}
A_0=&\ \frac{-2}{9(k^2 -1)^2 (2-2k^2 + \d_{ju}^2 +2\d_{rs}^2)(4k^2 -4 +\d_{ju}^2 +2\d_{rs}^2)^3} \Big\{
	4096 k(k^2-1)^6 
	+64\d_{rs}^2 (k^2-1)^5 ( 9 k^2 +56k +9)  
	\nonumber\\ &
	+96\d_{rs}^4 (k^2-1)^4 ( 9 k^2 -8k +9) 
	+16(k^2-1)^3 ( 27 k^2 -88 k +27 )  \d_{rs}^6 
	+8(k^2-1)^2 ( 9 k^2 -40k +9 )  \d_{rs}^8 
	\nonumber\\ &
	+\d_{ju}^2 \Big[ 32(k^2-1)^5 ( 9k^2 +14k +9) 
	+96\d_{rs}^2  (k^2-1)^4 ( 3 k^2 -35k +3 )
	-24\d_{rs}^4 (k^2-1)^3 ( 9 k^2 +166k +9  )
	\nonumber\\ &
	-8\d_{rs}^6  (k^2-1)^2 ( 36 k^2 +155k +36 )  
	-72 \d_{rs}^8 ( k^4 +k^3 -k -1 ) \Big]
	+\d_{ju}^4 \Big[ 48 k(k^2-1)^4 ( 6 k^2 +5 )  
	\nonumber\\ &
	+12\d_{rs}^2 (k^2-1)^3 ( 12k^3 -27k^2 -136k -27 ) 
	-6\d_{rs}^4 (k^2-1)^2 ( 36 k^3 +63k^2 +230k +63 )  
	\nonumber\\ &
	-36 \d_{rs}^6 ( 5k^5 +3k^4 
	+2k^3 -7k -3 )  
	-36\d_{rs}^8 k^3  \Big]
	-\d_{ju}^6 \Big[ 2(k^2-1)^3 ( 36k^3 +27k^2 + 394k +27) 
	\nonumber\\ &
	+2\d_{rs}^2 (k^2-1)^2 ( 108 k^3 +63k^2 +323k +63 )  
	+18 \d_{rs}^4 ( +9k^5 +3k^4 -4k^3 -5k -3 )  
	+36 \d_{rs}^6 k^3 \Big] 
	\nonumber\\ &
	-\d_{ju}^8 \Big[ (k^2-1)^2 ( 36k^3 +9k^2 -22k +9 )  
	+9 ( 4k^5 +k^4 -5k^3 +k -1 )  \d_{rs}^2 
	+9 k^3 \d_{rs}^4 \Big] 
	\Big\} \, ,
\end{align}
% different format of above equation
\begin{comment}
\begin{align}
A_0=&\ \frac{-2}{9(k^2 -1)^2 (2-2k^2 + \d_{ju}^2 +2\d_{rs}^2)(4k^2 -4 +\d_{ju}^2 +2\d_{rs}^2)^3} \bigg[
	4096 k(k^2-1)^6 
	\nonumber\\ &
		+64\d_{rs}^2 (k^2-1)^5 ( 9 k^2 +56k +9)  
		+96\d_{rs}^4 (k^2-1)^4 ( 9 k^2 -8k +9) 
	\nonumber\\ &\qquad
		+16(k^2-1)^3 ( 27 k^2 -88 k +27 )  \d_{rs}^6 
		+8(k^2-1)^2 ( 9 k^2 -40k +9 )  \d_{rs}^8 
	\nonumber\\ &
	+\d_{ju}^2 \Big( 32(k^2-1)^5 ( 9k^2 +14k +9) 
		+96\d_{rs}^2  (k^2-1)^4 ( 3 k^2 -35k +3 )
	\nonumber\\ &\qquad
	-24\d_{rs}^4 (k^2-1)^3 ( 9 k^2 +166k +9  )
		-8\d_{rs}^6  (k^2-1)^2 ( 36 k^2 +155k +36 )  
		-72 \d_{rs}^8 ( k^4 +k^3 -k -1 ) \Big)
	\nonumber\\ &
	+\d_{ju}^4 \Big( 48 k(k^2-1)^4 ( 6 k^2 +5 )  
		+12\d_{rs}^2 (k^2-1)^3 ( 12k^3 -27k^2 -136k -27 ) 
	\nonumber\\ &\qquad
		-6\d_{rs}^4 (k^2-1)^2 ( 36 k^3 +63k^2 +230k +63 )  
		-36 \d_{rs}^6 ( 5k^5 +3k^4 +2k^3 -7k -3 )  
		-36\d_{rs}^8 k^3  \Big)
	\nonumber\\ &
	-\d_{ju}^6 \Big( 2(k^2-1)^3 ( 36k^3 +27k^2 + 394k +27) 
		+2\d_{rs}^2 (k^2-1)^2 ( 108 k^3 +63k^2 +323k +63 )  
	\nonumber\\ &\qquad
		+18 \d_{rs}^4 ( +9k^5 +3k^4 -4k^3 -5k -3 )  
		+36 \d_{rs}^6 k^3 \Big) 
	\nonumber\\ &
	-\d_{ju}^8 \Big( (k^2-1)^2 ( 36k^3 +9k^2 -22k +9 )  
		+9 ( 4k^5 +k^4 -5k^3 +k -1 )  \d_{rs}^2 
		+9 k^3 \d_{rs}^4 \Big) 
	\bigg] \, ,
\end{align}
\end{comment}

\begin{align}
A_{tan} =&\ \frac{4k}{(k-1) (2 -2k^2 +\d_{ju}^2 +2\d_{rs}^2)^2 \sqrt{(k-1)^3(k+1)}}\, 
	\textrm{arctan} \bigg( \frac{\sqrt{(k-1)^3 (k+1)}}{k^2+k -1} \bigg)
	\Big\{
	8(k-1)^4 (k+1)^3  
	+4\d_{rs}^2 (k^2 -1)^2 
	\nonumber\\ &
	+4\d_{rs}^4 ( k^3 -2k^2 -k +2 )
	-\d_{ju}^2 \Big[ 8(k-1)^3 (k+1)^2 +4\d_{rs}^2 (k^2-1) -2 \d_{rs}^4 \Big]
	+\d_{ju}^4 \Big[ 2(k-1)^2 (k+1) +\d_{rs}^2 \Big] 
	\Big\}
	\nonumber\\ \nonumber\\ &
	-\frac{8k (2 -2k^2 +\d_{ju}^2 -\d_{rs}^2)^2 
		\sqrt{(8k^2 -12k +4 -\d_{ju}^2 -2\d_{rs}^2)(4k^2 -4 +\d_{ju}^2 +2\d_{rs}^2)}}
		{9(k-1)^2 (2-2k^2 +\d_{ju}^2 +2\d_{rs}^2)^2}
	\nonumber\\ &
	\times \textrm{arctan} \bigg( \frac{\sqrt{(8k^2 -12k +4 -\d_{ju}^2 -2\d_{rs}^2)(4k^2 -4 +\d_{ju}^2 +2\d_{rs}^2)}}{4k^2 +6k -4 +\d_{ju}^2 +2\d_{rs}^2} \bigg)\, .
\end{align}

% different format of above equation
\begin{comment}
\begin{align}
A_{tan} =&\ \frac{4k}{(k-1) (2 -2k^2 +\d_{ju}^2 +2\d_{rs}^2)^2 \sqrt{(k-1)^3(k+1)}}\, 
	\textrm{arctan} \left( \frac{\sqrt{(k-1)^3 (k+1)}}{k^2+k -1} \right)
	\nonumber\\ &\qquad
	\times \bigg[
		8(k-1)^4 (k+1)^3  
		+4\d_{rs}^2 (k^2 -1)^2 
		+4\d_{rs}^4 ( k^3 -2k^2 -k +2 )
	\nonumber\\ &\qquad\qquad
		-\d_{ju}^2 \Big( 8(k-1)^3 (k+1)^2 +4\d_{rs}^2 (k^2-1) -2 \d_{rs}^4 \Big)
		+\d_{ju}^4 \Big( 2(k-1)^2 (k+1) +\d_{rs}^2 \Big) 
	\bigg]
	\nonumber\\ \nonumber\\ &
	-\frac{8k (2 -2k^2 +\d_{ju}^2 -\d_{rs}^2)^2 
		\sqrt{(8k^2 -12k +4 -\d_{ju}^2 -2\d_{rs}^2)(4k^2 -4 +\d_{ju}^2 +2\d_{rs}^2)}}
		{9(k-1)^2 (2-2k^2 +\d_{ju}^2 +2\d_{rs}^2)^2}
	\nonumber\\ &
	\times \textrm{arctan} \left( \frac{\sqrt{(8k^2 -12k +4 -\d_{ju}^2 -2\d_{rs}^2)(4k^2 -4 +\d_{ju}^2 +2\d_{rs}^2)}}{4k^2 +6k -4 +\d_{ju}^2 +2\d_{rs}^2} \right)\, .
\end{align}
\end{comment}

\end{widetext}

\bibliography{%
bib_files/general,%
bib_files/EFT,%
bib_files/lattice_PQ,%
bib_files/lattice_FV,%
bib_files/NNEFT,%
bib_files/lattice_fermions,%
bib_files/lattice_physics,%
bib_files/lattice_general,%
bib_files/heavy_mesons,%
bib_files/hadron_structure}

\end{document}